\documentclass[12pt,
 superscriptaddress, 
 notitlepage,
 longbibliography,
 notitlepage]{revtex4-1}
 
\usepackage{graphicx}
\usepackage{verbatim}
\usepackage{todonotes}
\usepackage{amsmath}
\usepackage{color,soul}
\usepackage{amsmath,mleftright}
\usepackage{xparse}
\usepackage{subfiles}
\usepackage{float}
\usepackage{subfig}
\usepackage[nomargin,inline,marginclue,draft]{fixme}
\usepackage{caption}

\usepackage{enumitem, hyperref}

\makeatletter

\renewcommand*\FXLayoutInline[3]{%
  {\@fxuseface{inline}\ignorespaces[#3 \fxnotename{#1}: #2]}}
\makeatother

\NewDocumentCommand{\evalat}{sO{\big}mm}{%
  \IfBooleanTF{#1}
   {\mleft. #3 \mright|_{#4}}
   {#3#2|_{#4}}% 
   }

\begin{document}

\title{Effect of proximity to ionic liquid-solvent demixing on electrical double layers}

\author{Carolina Cruz}
\author{Svyatoslav Kondrat}
%\email{svyatoslav.kondrat@gmail.com}
%\email{skondrat@ichf.edu.pl}
\affiliation{Department of Complex Systems, Institute of Physical Chemistry PAS,  Warsaw, Poland}

\author{Enrique Lomba}
%\affiliation{Institute of Physical Chemsitry Rocasolano,  CSIC, Madrid, Spain}
\affiliation{Instituto de Qu{\'\i}mica F{\'\i}sica Rocasolano, CSIC, Serrano 119, E-28006 Madrid, Spain}

\author{Alina Ciach}
\affiliation{Department of Complex Systems, Institute of Physical Chemistry PAS,
 Warsaw, Poland}

 \date{\today}
 
\begin{abstract}
    There is a growing interest in the properties of ionic liquids (ILs) and IL-solvent mixtures at metallic interfaces, particularly due to their applications in energy storage. The main focus so far has been on electrical double layers with ILs far from phase transitions. However, the systems in the vicinity of their phase transformations are known to exhibit some remarkable features, such as wetting transitions and capillary condensation. Herein, we develop a mean-field model suitable for the IL-solvent mixtures close to demixing, and combine it with the Carnahan-Starling (CS) and lattice-gas expressions for the excluded volume interactions. This model is then solved analytically, using perturbation expansion, and numerically. We demonstrate that, besides the well-known camel and bell-shaped capacitances, there is a bird-shaped capacitance, having three peaks as a function of voltage, which emerge due to the proximity to demixing. In addition, we find that the camel-shaped capacitance, which is a signature of dilute electrolytes, can appear at high IL densities for ionophobic electrodes. We also discuss the differences and implications arising from the CS and lattice-gas expressions in the context of our model.

\end{abstract}

 \maketitle
 
\section{Introduction}

Ionic liquids (ILs) and IL--solvent mixtures have become the focus of research in electrochemistry due to their unique properties, such as exceptional electrochemical and thermal stability, and low vapour pressure. This makes them attractive materials for many applications \cite{Welton1999, Fedorov2014}, for instance, for electrochemical reactions, as lubricants for micro and nanodevices \cite{C2CP23814D}, as extraction liquids for the purification of metals, colloids and biomass, etc \cite{Fedorov2014}. The classical theory of electrolytes was developed in the early $20th$ century, with the achievements of Gouy, Chapman, Debye, H\"{u}ckel and Langmuir channeled into the so-called Poisson-Boltzmann (PB) model \cite{Ben-Yaakov2011}. The PB model is a mean-field model which describes ions as isolated point-like charges in a solvent considered as a continuum dielectric \cite{Parsons1990}. Electrical double layers (EDLs) emerge when electrolytes are put in contact with charged surfaces. An EDL results from the formation of an ion `cloud' of opposite sign to that of the surface charge, and its width is influenced by the competition between the thermal motion of the ions, which tends to homogenize their distribution, and the Coulomb interactions, which attract the counterions to  the surface \cite{Parsons1990}. Within the linearized PB (Debye--H\"{u}ckel approximation), the thickness of this layer is given by the Debye length $\lambda_D=(4\pi\rho_b\lambda_B)^{-1/2}$, where $\rho_b$ is the ion density and $\lambda_B$ the Bjerrum length. The application of the PB model to EDLs, known as the Gouy--Chapman model, predicts the well-known U-shaped dependence of the EDL capacitance on the applied potential.

However, the classical description of ELDs is only valid for dilute electrolytes (concentrations below $0.01$M), but it is not suitable for ILs due to typically high concentrations of ions, at which the ion sizes start to play a role \underline{\cite{Bikerman1942,wicke1952einfluss,Freise1952,Eigen1954}}. Indeed, theories developed for EDLs have shown that excluded volume interactions are crucial to describe the structure of the EDL with ILs properly \cite{PhysRevLett.79.435, DiCaprio2003, antypov05a, Oldham2008, McEldrew2018,Bohinc2001a,Kornyshev2007a, kilic:pre:07a, Frydel2012, Minton2016, Girotto2018}. Steric interactions restrict the absorption of counterions at an electrode, and hence influence the charge density in the EDL. This leads to the emergence of the so-called camel and bell-shape capacitances, obtained at low and high IL concentrations, respectively \cite{Kornyshev2007a}, instead of the classical Gouy-Chapman's U-shape.

Temperature also plays an important role in the structure and capacitance of EDLs. However, contradictory results have been reported in the literature and consensus is yet to be reached as to whether capacitance increases or decreases with temperature and under which conditions. According to the Gouy-Chapman theory, the capacitance decreases for increasing temperature, but the experiments showed also the opposite trends \cite{Lockett2008b, Silva2008,Druschler2012a,Ivanistsev2017a}. For instance, \citeauthor{Silva2008} \cite{Silva2008} studied the [BMIM][PF6] ionic liquid and three different electrodes and found that the differential capacitance increases with temperature at all potentials.  \citeauthor{Lockett2008b} \cite{Lockett2008b} found the same behaviour for imidazolium-based ionic liquids in contact with glassy carbon electrodes. More careful theoretical work suggested that both trends are possible
\cite{holovko:01,Reszko-Zygmunt2005,chen_kornyshev:18}, but there is no general agreement on the origin of this behaviour. For instance, \citeauthor{holovko:01} \cite{holovko:01} proposed that the increase of capacitance is related to the decreased inter-ionic interactions and weaker ion associations, while \citeauthor{chen_kornyshev:18} \cite{chen_kornyshev:18} argued that the temperature dependence of EDLs is chiefly determined by the strength and extent of the van der Waals interactions. Interestingly, it was shown that the transition between the camel and bell-like capacitances can be induced also by varying temperature \cite{Vatamanu2011a,chen_kornyshev:18}, with the bell shape emerging at high temperatures due to breaking of `ion pairs' and consequently stronger screening \cite{chen_kornyshev:18}.

The focus of the above-mentioned studies was on room--temperature ILs and IL--solvent mixtures far from phase transitions. However, it is well--known that, for neutral fluids, the fluid structure at a surface undergoes drastic changes  (such as wetting or prewet-ting transitions, layering, etc), when the system approaches a phase transition \cite{Bonn2009}. In a recent work \cite{Cruz2019a}, we have proposed a model suitable for IL--solvent mixtures close to demixing. Treating solvent as a continuum (appropriate for small solvent molecules, such as water or acetonitrile), but describing ILs as amenable to phase separate into the ion-rich and ion-dilute phases, we showed that the capacitance and stored energy become sensitive functions of temperature in the vicinity of demixing. We also demonstrated the emergence of a new, \emph{bird}-shaped capacitance, having three peaks as a function of voltage. Herein, we present a more detailed investigation of this system, while we also extend our study to systems, in which solvent and ions are of comparable size. We first describe the details and the derivation of the model (Sec.~\ref{theory}), and discuss its phase behaviour in bulk (Sec.~\ref{Bulk phase diagram}). Then, we derive analytic expressions for the density, potential and charge profiles, as well as for the capacitance, by applying the perturbation expansion (Sec.~\ref{Sec:Analytical_solution}). The results of numerical calculations are discussed in Sec.~\ref{Section:Num_Results}. We summarize in Sec.~\ref{Section:Conclusion}.

 \section{Model}
 \label{theory}
 
 We consider a mixture of ionic liquid (IL) and  neutral solvent, which can phase separate below its upper critical point \cite{Elshwishin2014}. Our interest is in the one-phase region just above demixing. We assume that the mixture is in contact with a planar metallic electrode, and the electrostatic potential, $U$, is kept constant with respect to the bulk. This system can be described by the following grand thermodynamic potential 
 \begin{equation}
 \label{om}
\Omega[\rho_{\pm},u]/A = \omega_{el}+ \omega_{vdW}-Ts-\mu_i\int_0^{\infty}dz \rho_i(z),
 \end{equation}
where $\omega_{el}$, $\omega_{vdW}$ and $s$ are the electrostatic energy, the energy associated with van der Waals-like dispersion (non-Coulombic) interactions and the entropy (all per surface area), respectively; $T$ is temperature, $A$ is the surface area of the electrode, $\mu_i$ is the chemical potential, and $\rho_i(z)$ is the local density of different components, where $i= \left\lbrace +,-,s\right\rbrace$ denotes cations, anions and solvent, respectively. The electrostatic energy is given by \underline{\cite{PhysRevLett.79.435, Kralj-Igli1996}} 
 \begin{equation}
 \label{uel}
  \beta\;\omega_{el}[c(z), u(z)]=\int_0^{\infty} dz \Big[cu-\frac{1}{8\pi \lambda_B}\Big (\frac{\partial u}{\partial z}\Big)^2\Big],
 \end{equation}
 where $\beta=1/(k_B T)$, $u$ is the electrostatic potential in $k_BT/e$ units with $k_B$ denoting the Boltzmann constant; $c=\rho_+-\rho_-$ is the charge density per elementary charge $e$ and $\lambda_B=\beta e^2/\epsilon$ is the Bjerrum length, where $\epsilon$ is the dielectric constant. It is well known that $\epsilon$ depends on temperature, particularly for polar solvents \cite{Gagliardi2007,Riniker2012,Orhan2014}. Nevertheless, we assume $\epsilon$ to be temperature-independent, and note that its temperature variation should not affect the results qualitatively, as pointed out in \cite{Cruz2019a}. In addition, it is known that polarizability of solvent and of ions may play an important role in the structure and properties of electrical double layers \cite{Gongadze2012, Fedorov2008b, Fedorov2010, Girotto2017, Pousaneh2012a, Pousaneh2014}. In particular, \citeauthor{Gongadze2012} \cite{Gongadze2012} demonstrated a potentially strong variation of the dielectric constant close to a planar charged surface. These authors derived a formula for a position-dependent $\epsilon$, by taking into account excluded volume effects and solvent polarization. In general, such variation of $\epsilon$ shall also depend on the affinity of ions/solvent towards electrode (electrode's ionophilicity, see below), as well as on the applied potential. To avoid such complications, and to capture generic effects, unobscured by the chemical complexity, we have decided to take a position-independent dielectric constant, as in the majority of studies \cite{PhysRevLett.79.435, DiCaprio2003, Kornyshev2007a, Fedorov2008a, Oldham2008, Fedorov2010, Girotto2017, Girotto2018, chen_kornyshev:18, McEldrew2018}; clearly, the change of $\epsilon$ close to the surface will affect the results quantitatively, but it is reasonable to expect that the qualitative behaviour will not be altered.

 The term $\omega_{vdW}$ describes the contribution from attractive non-Coulombic van der Waals-like interactions to the internal energy, which may lead to demixing of the IL and solvent. The explicit expression for $\omega_{vdW}$ is difficult to obtain due to the complexity of the interactions between the ions and solvent. However, when the phase separation is driven by the chemical difference between IL and neutral solvent, we can take into account only the effective interactions leading to phase separation. In this case, for the bulk system one can write
 \begin{equation}
 \label{USR3}
 \omega_{vdW}-\sum_i\mu_i\int d{\bf r}\rho_i({\bf r})=
 \frac{1}{2}\int d{\bf r}_1\int  d{\bf r}_2 J(r)g(r)\rho({\bf r}_1)\rho({\bf r}_2)
 -\mu\int d{\bf r}\rho({\bf r}) ,
\end{equation}
where $\mu$ is the difference between the chemical potentials of an IL and solvent, $J(r)$ represents the effective interactions leading to phase separation, $g(r)$ is the pair distribution function, $r=|{\bf r}_1-{\bf r}_2|$, and $\rho({\bf r})=\rho_+({\bf r})+\rho_-({\bf r})$ is the number density of the IL  at position ${\bf r}$. We have used the approach of Ref.~\cite{Pousaneh2012a,Pousaneh2014} to transform Eq.~\eqref{USR3} to
\begin{equation}
 \label{USR4}
 \Big[\omega_{vdW}-\sum_i\mu_i\int d{\bf r}\rho_i({\bf r})\Big]/A
 \approx K\Bigg\{
 \int_0^{\infty}dz \Bigg[\frac{\xi_0^2}{2} \Bigg(\frac{\partial \rho}{\partial z}\Bigg)^2-\frac{1}{2}\rho^2 \Bigg] + \frac{\xi_0}{2} \rho^2(0) - h_1 \rho(0) \Bigg\} -\mu\int_0^{\infty} dz\rho,
\end{equation}
where 
\begin{equation}
 \label{J}
 K=- \int d{\bf r}J(r)g(r)>0
\end{equation} 
measures the strength of the dispersion interactions, and 
\begin{equation}
 \label{xi0}
 \xi_0^2=\frac{1}{2}\frac{\int d{\bf r}J(r)g(r)r^2}{\int d{\bf r}J(r)g(r)}
\end{equation}
describes the spatial extension of these interactions ($\xi_0$ is of the same order of magnitude as the molecular size $a$). In Eq.~\eqref{USR4}, we took into account that the interactions with the missing fluid neighbours beyond the system boundary should be subtracted (the first boundary term), and we included the direct short-range interactions of the fluid particles with the wall (the second boundary term). The electrode's ionophilicity is denoted by $h_1$ and describes the preference of the electrode for ions or solvent; $h_1 > 0$ means that the wall favours ions, and we assumed this preference to be the same for anions and cations. 

Within the local density approximation, the entropy is
 \begin{equation}
 \label{Ts}
  -Ts =-T\int_0^{\infty} dz s([\rho_i(z)])= k_BT \int_0^{\infty} dz \Bigg[\rho_+\ln(a^3 \rho_+)+\rho_-\ln(a^3 \rho_-)
 +\beta f_{ex}\Bigg].
 \end{equation}
The first two terms in Eq.~\eqref{Ts} come from the entropy of mixing of ions, and the last term is the excess free energy associated with the excluded volume interactions. If the cations and anions are of comparable size, but the solvent molecules are much smaller, such that the solvent can be treated as a structureless continuum, it seems reasonable to use the Carnahan - Starling (CS) approximation \cite{Carnahan1969EquationSpheres} for the excluded volume interactions between the ions only, \textit{i.e.}, 
 \begin{eqnarray}
 \label{fexCS}
 \beta f_{ex}^{CS}(\rho)=\rho\Bigg(
 \frac{4\eta -3\eta^2}{(1-\eta)^2}-1
 \Bigg),
\end{eqnarray}
where $\eta=\pi \rho a^3 /6$ is the packing fraction of ions. However, if both ions and solvent are of comparable size, it might be more suitable to use the popular lattice-gas expression  
\begin{align}
  \label{fexBK}
   \beta f_{ex}^{lg}(\rho)=(\rho_{tot}-\rho)\ln \big[a^3(\rho_{tot}-\rho)\big],
 \end{align}
which arises from the solvent's ideal-gas entropy, $ \beta f_{ex}=\rho_s\ln a^3 \rho_s$, by assuming the local incompressibility conditions, $\rho_+({\bf r})+\rho_-({\bf r})+\rho_s({\bf r})=\rho_{tot}$ ($/rho_{tot}=a^{-3}$ for the lattice--gas model). Eq.~\eqref{fexBK} has been employed in a number of important studies, most notably by \citeauthor{Bikerman1942} \cite{Bikerman1942}, \citeauthor{wicke1952einfluss} \cite{wicke1952einfluss, Eigen1954}, \citeauthor{PhysRevLett.79.435} \cite{PhysRevLett.79.435}, \citeauthor{kilic:pre:07a} \cite{kilic:pre:07a} and \citeauthor{Kornyshev2007a} \cite{Kornyshev2007a}.

In Eqs.~\eqref{fexCS} and \eqref{fexBK}, the cations and anions are assumed to be of the same size, i.e., $a_-=a_+=a$, whereas often $a_{+}\neq a_{-}$~\cite{Gongadze2015, Sin2015, Gongadze2018}. \citeauthor{Gongadze2015}~\cite{Gongadze2015} proposed an improved mean-field model of EDLs that accounts for such ion-size asymmetry, and found that it leads to a pronounced decrease of the capacitance and to shape asymmetry of the capacitance-voltage curves (with respect to the potential of zero charge), which seems to be consistent with the experimental observations~\cite{Lockett2008c,Lockett2010}. These results suggest that the asymmetry in ion sizes may reduce the capacitance calculated in this work, and will  additionally bring asymmetry in the capacitance-voltage dependence, but the qualitative behaviour due to proximity to demixing shall be captured already by a model featuring the same sizes of cations and anions.

Summing up, our final expression of the grand potential is \cite{Cruz2019a}
\begin{multline}
\label{Eq_GP}
\beta\; \Omega[\rho_{\pm},u]/A = 
    \int_0^{\infty} dz \Bigg[ \rho_{+} \ln(a^3 \rho_{+})+\rho_{-} \ln(a^3 \rho_{-})+\beta f_{ex}(\rho)\Bigg]
    + \int_0^{\infty} dz \Bigg[ cu - \frac{1}{8 \pi \lambda_B} \Bigg( \frac{\partial u}{\partial z}\Bigg)^2\Bigg] \\
    + \beta K \left\lbrace \int_0^{\infty} dz \Bigg[ \frac{\xi_0^2}{2} \Bigg( \frac{\partial \rho}{\partial z}\Bigg)^2- \frac{1}{2}\rho^2 \Bigg] + \frac{\xi_0}{2}\rho_0^2 - h_1 \rho_0\right\rbrace
    - \beta \mu \int_0^{\infty} \rho dz.
\end{multline}

The equilibrium properties of the system are described by the minimum of $\Omega$. Minimization with respect to $u$ and $c$ yields
\begin{equation}
%\label{ELf1}
    \lambda_D^2 u'' = - c = (1 + \phi/\bar{\rho_b}) \tanh(u),
    \label{eq:potential}
\end{equation}
where $\lambda_D=(4\pi\rho_b\lambda_B)^{-1/2}$ is the Debye screening length in bulk electrolyte and $\rho_b$ is the equilibrium ion density (in bulk) and $\phi=\bar \rho -\bar \rho_{b}(\bar \rho_{b} = a^3 \rho_b)$. The boundary conditions are, naturally, $u(\infty)=0$ and $u(0)= eU/k_BT$, where $U$ is the potential applied at an electrode with respect to bulk. Minimization with respect to $\rho$ gives
\begin{equation}
%\label{ELf2}
    \xi_0^2\; \phi'' + \phi = \bar{T} \left[ \ln (1 + \phi/\bar{\rho_b})- \ln(\cosh(u)) + \Delta \mu_{ex} \right],
    \label{eq:iondensity}
\end{equation}
 where $\Delta \mu_{ex} = \mu_{ex}-\mu^{b}_{ex}$ with $\mu_{ex}=\beta \partial f_{ex}/\partial \rho$, $\mu^{b}_{ex}=\mu_{ex}(z=\infty)$, and $\bar{T}= k_B T a^3/K$ is dimensionless temperature. The boundary conditions are $ \phi(\infty) = 0$ and $ \xi_0 \phi'(0) - \phi(0) + \tilde{h_1} = 0$, where $\tilde{h_1}= a^3h_1/\xi_0-\rho_b$. From the solution of Eq.~\eqref{eq:potential}-\eqref{eq:iondensity}, the total charge, $Q$, stored in an EDL is
\begin{equation}
 \label{Eq:Q}
 Q=-e\int_0^{\infty}c dz = \evalat[\bigg]{\frac{4 \pi}{\epsilon} \frac{du}{dz}}{z=0}.
\end{equation}

An important quantity, which can be assessed experimentally, is the differential capacitance: 
\begin{equation}
    C=\frac{\partial Q}{\partial U}.
    \label{Eq:DiffCap}
\end{equation}

We have calculated these quantities both numerically and analytically, and the results are discussed in Sections~\ref{Sec:Analytical_solution} and \ref{Section:Num_Results}. First, however, we briefly describe the bulk system, \textit{i.e.}, the system in the absence of an electrode.

\section{Bulk phase diagram}
\label{Bulk phase diagram}

\begin{figure}[t]
\centering
\includegraphics[scale=0.8]{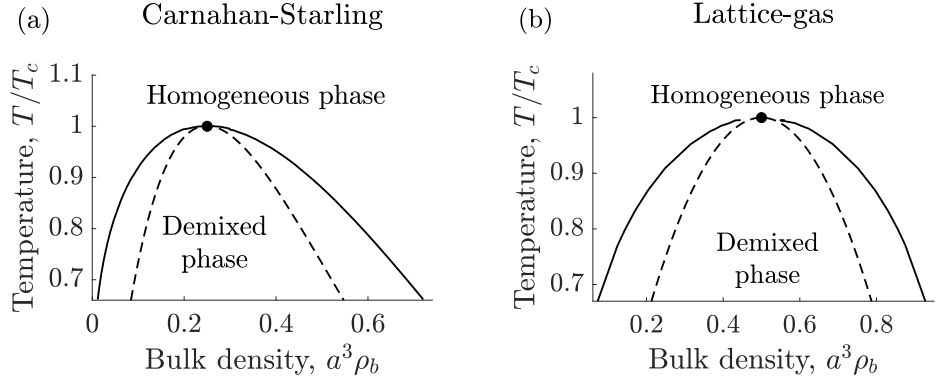}
\label{fig:foo-1}
\caption{\textbf{Bulk phase diagrams of ionic liquid (IL)-solvent mixtures.} The solid lines represent the first order phase transition between the homogeneous and IL-solvent demixed phases, and the dashed lines are the spinodal curves given by Eq. \eqref{eq:spinodal}. The circles denote (upper) critical points. Temperature is expressed in terms of the critical temperature $T_c$.}
\label{Phase_Diagrams}
\end{figure}

In the absence of electrode, the thermodynamic potential is given by
\begin{equation}
\label{Omegab}
 \beta \; \Omega_{b}(\rho_b)/A = -\beta K \rho_b^2 /2 + \Big[
 \rho_b\ln(\rho_b/2 )+ \beta f_{ex}(\rho_b)\Big]-\mu\rho_b.
\end{equation}
The equilibrium condition, $\partial\Omega_b/\partial\rho_b=0$, leads to a non-linear equation, which we solved numerically. Our analysis indicates that, in some parameter region, there are two solutions, corresponding to the IL-rich ($\rho_b=\rho_1$) and IL-poor ($\rho_b=\rho_2$) phases. The first order phase transition between these phases occurs when $\Omega_{b}(\rho_1) = \Omega_{b}(\rho_2)$. This is shown in Fig.~\ref{Phase_Diagrams} by solid lines for the CS and lattice-gas models.

At the spinodal line, $\delta^2\Omega_b/\delta \rho_b^2=0$, the homogeneous IL-solvent mixture becomes unstable with respect to density fluctuations (this corresponds to the diverging correlation length, c.f. Eq.~\eqref{xi}). Within our mean-field  theory, the spinodal line is given by 
\begin{equation}
\label{eq:spinodal}
 \bar{T}_c(\bar{\rho}_b)=\alpha^{-1}(\bar{\rho}_b),
\end{equation}
where
\begin{equation}
\label{w}
 \alpha (\bar{\rho}_b)=\Bigg[\evalat[\bigg]{\frac{\partial \mu_{ex}}{\partial \bar{\rho}}}{\bar{\rho}=\bar{\rho}_b}+\bar{\rho}_b^{-1}
 \Bigg].
\end{equation}
The spinodals are shown by dashed lines in Fig.~\ref{Phase_Diagrams}. The  point on the spinodal that satisfies $d\bar{T}_c(\bar{\rho}_b)/d\bar{\rho}_b=0$ corresponds to a critical point (solid circles in the same figure). For the CS and lattice-gas expressions, we found for the critical points $\bar{\rho}_c\approx0.25$, $\bar{T}_c\approx0.09$ and $\bar{\rho}_c\approx0.5$, $\bar{T}_c\approx 0.25$, respectively. 

We note that the obtained phase diagrams (Fig.~\ref{Phase_Diagrams}) are in good qualitative agreement with the experimental data (see, \textit{e.g.}, Refs.~\cite{butka:08:ILPhaseTransitions, Crosthwaite2004, rotrekl_bendova:17:ILSDemixing}).

\section{Approximate analytical solution}
\label{Sec:Analytical_solution}

To study the behaviour of IL-solvent mixtures at metallic surfaces within our model, one needs to solve Eqs. \eqref{eq:potential} - \eqref{eq:iondensity}, which we have done numerically. However, before describing the results of those calculations, it is useful to discuss approximate analytical solutions, which can be obtained for weak surface potentials, $U$, and ionophilicities, $\tilde{h}_1$. To this end, we used the standard perturbation analysis, \textit{i.e.}, we assumed $u=u_0+\varepsilon u_1+ \varepsilon^2 u_2+\varepsilon^3 u_3...$ and $\phi = \phi_0+\varepsilon \phi_1+ \varepsilon^2 \phi_2+... $, where $\varepsilon$ is a small parameter. In the first  order approximation, we obtained the following equations
\begin{align}
 \frac{d^2u_1}{dz^2} =\lambda_D^{-2} u_1(z)
    \label{ELl1}    
\end{align}
and
\begin{align}
    \frac{d^2\phi_1}{dz^2} =\xi^{-2}\phi_1(z),
    \label{ELl2}    
\end{align}
where 
\begin{equation}
\label{xi}
% \xi=\xi_0\big(\bar{T}\alpha (\bar{\rho}_b)-1\Big)^{-1/2}
    \xi = \xi_0 \Big( \frac{\bar T}{\bar T_c (\bar \rho_b)}-1 \Big)^{-1/2}
\end{equation}
is the correlation length and $\bar T_c$ is given by Eq. \eqref{eq:spinodal}. The solutions to Eqs.~\eqref{ELl1} and \eqref{ELl2} are  
\begin{equation}
    u_{1}(z) =U e^{-\kappa z}
\end{equation}
and
\begin{equation}
 \phi_{1}(z) = \frac{\tilde{h_1}}{\xi_0/\xi+1}e^{-z/\xi}. 
\end{equation}
Thus, as one may expect, in the first order approximation, the fields $u$ and $\phi$ are fully decoupled, that is, the behaviour of $u$ is determined solely by the Debye screening length, $\lambda_D$, as in the classical Debye-H\"uckel theory, while the decay of $\phi$ is governed by the correlation length, $\xi$.

\begin{figure}[t]
\centering
\includegraphics[scale=0.9]{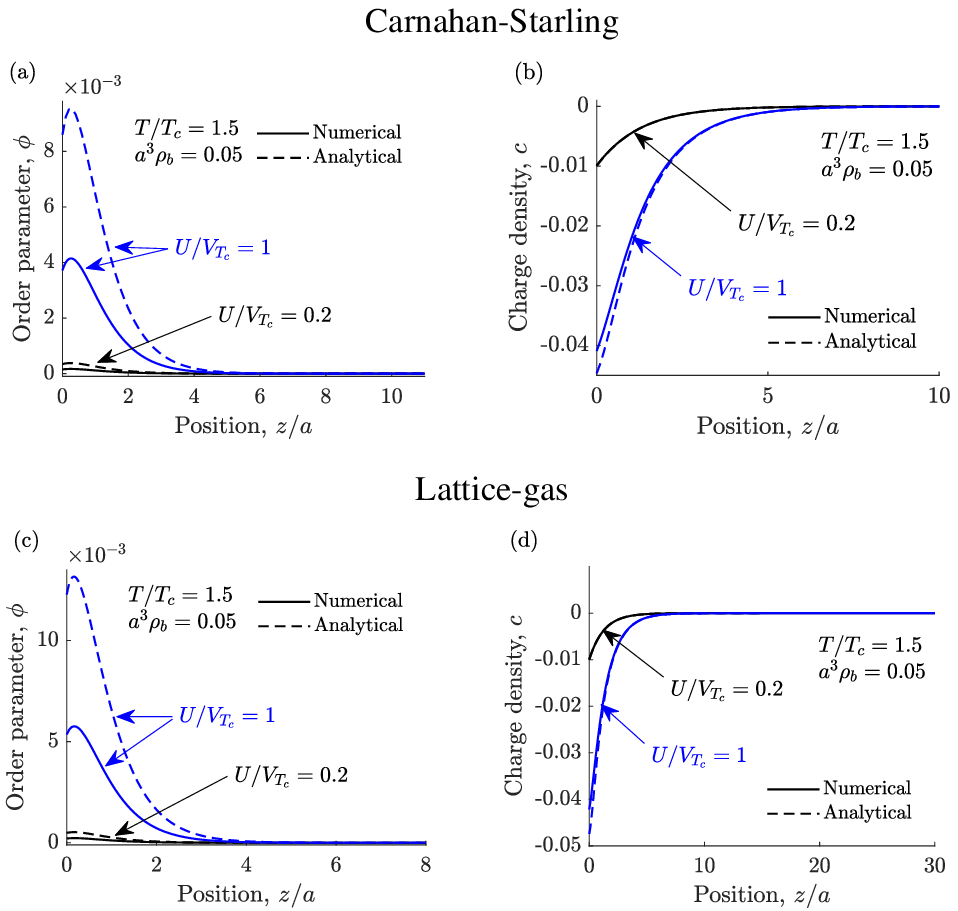}
    \caption{\textbf{Density profiles for the Carnahan-Starling and lattice-gas entropies.} The dashed and solid lines correspond to the analytical  and numerical solutions, respectively. The profiles were obtained at the applied potentials, $U$, as indicated on the plots, for ionophilicity $\tilde{h}_1=a^3h_1/\xi_0-\bar \rho_b=0$, and for $\xi_0=\lambda_B=a$. (\textbf{a}) Order parameter $\phi=a^3\;(\rho-\rho_b)$ for the Carnahan-Starling (CS) entropy, where $\rho=\rho_{+}+\rho_{-}$ is the total ion density. (\textbf{b}) Charge density $c=a^3\;(\rho_{+}-\rho_{-})$ for the CS approximation. (\textbf{c}) and (\textbf{d}) The same as (a) and (b) but for the lattice-gas entropy. Voltage is expressed in terms of the thermal voltage $V_{T_c}= k_B\;T_c/e$ taken at $T_c$.}
\label{Profiles}
\end{figure}

In the second order perturbation, we obtained
\begin{equation}
\label{eq:u2}
 \frac{d^2 u_2}{dz^2}  = \lambda_D^{-2} u_2 + \frac{\lambda_D^{-2}}{\bar{\rho}_b}\phi_1 u_1   
\end{equation}
and
\begin{equation}
\label{eq:phi2}
 \frac{d^2 \phi_2}{dz^2} = \xi^{-2} \phi_2 + \frac{\bar{T}}{2 \; \xi_0^2 \; \bar{\rho}_b^2} \Bigg[A(\bar \rho_b) \phi_1^2- \bar{\rho}_b^2 u_1^2\Bigg],
\end{equation}
 where    
\begin{equation}
    A(\bar{\rho}_b) = \bar{\rho}_b^2 \; \evalat[\bigg]{\frac{\partial^2 \bar{\mu}_{ex}}{\partial \bar{\rho}^2}}{\bar{\rho}=\bar{\rho}_b} -1.
\end{equation}
Note that $A(\bar{\rho}_b=\bar{\rho}_c)=0$, where $\bar{\rho}_c$ is the critical density. Thus, in the second order, the $u$ and $\phi$ fields become coupled, and it is this coupling that determines the highly non-linear behaviour of the system as it approaches demixing. The solutions to Eqs.~\eqref{eq:phi2} and \eqref{eq:u2} are lengthy and are not presented here.

Fig.~\ref{Profiles} compares the second--order analytical and numerical solutions to Eqs.~\eqref{eq:potential} and \eqref{eq:iondensity} for the Carnahan-Starling and lattice-gas entropies. For the order parameter, $\phi$, analytic and numerical solutions differ significantly for increasing the applied potential. Interestingly, however, for the charge density the perturbation expansion provides a relatively good approximation and the solutions agree even for higher potentials.

\subsection{Differential capacitance}

Differential capacitance can be computed by plugging the electrostatic potential, $u$, obtained by the perturbation expansion, into Eq.~\eqref{Eq:Q} and \eqref{Eq:DiffCap}; the result is 
\begin{equation}
 \label{Cper}
 C = C_0+ C_2 U^2+...,
\end{equation}
where
\begin{equation}
\label{C0}
C_0= C_{D} \Big[ 1 + \frac{\tilde h_1\;\xi/\lambda_D}{(1+\xi_0/\xi)(2\xi/\lambda_D +1)} + O(\tilde h_1^2)\Big]    
\end{equation}
and
\begin{equation}
%\label{C2}
 C_2=\frac{C_{D}}{4}\Bigg[ \frac{3(4\xi/\lambda_D +\xi/\xi_0 +1)\big[(\xi/\xi_0)^{2}+1\big]}
 {2 (\xi/\xi_0 +1)(2\xi/\lambda_D +1)^2 \bar{\rho}_b \; \alpha(\bar{\rho}_b)}
 -1\Bigg]+O(\tilde h_1),
 \label{eq:C2_big}
\end{equation}
where $C_{D}=(a/\lambda_D) C_H$ is the Debye capacitance and $C_H=\epsilon/4 \pi a$ the Helmholtz capacitance. 

The sign of $C_2$ describes the shape of the capacitance at low potentials. A positive $C_2$ corresponds to the so-called camel shape, exhibiting a minimum at $u=0$, while a negative $C_2$ means a maximum at $u=0$ and is often associated with the bell-shaped capacitance \cite{Kornyshev2007a}. Such capacitance shapes have been extensively studied in the literature \cite{Kornyshev2007a, Lockett2008b, chen_kornyshev:18,Vatamanu2011a,Fedorov2008a,Alam2008a,Fedorov2008b,Fedorov2010, Girotto2017, Girotto2018}. 

\begin{figure}[t]
\centering
\includegraphics[scale=0.9]{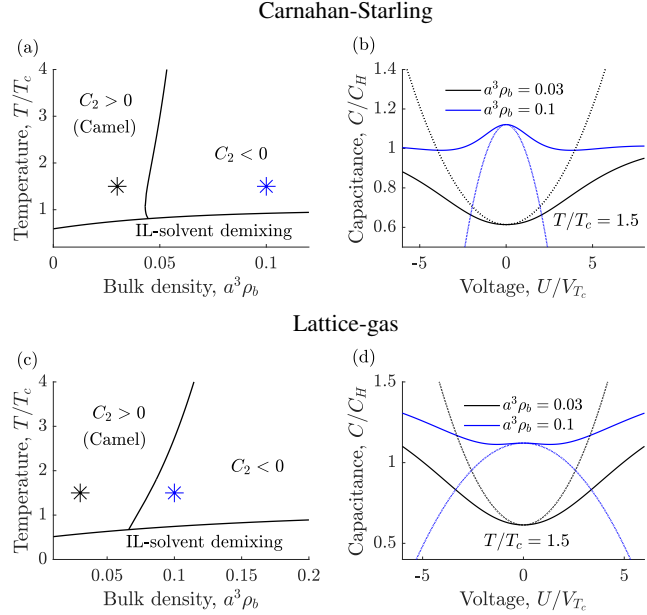}
\caption{\textbf{Capacitance close to demixing from perturbation expansion.} \textbf{(a)} Capacitance diagram showing the region of positive and negative curvature in the low-voltage capacitance ($C_2>0$ and $C_2<0$, respectively, see Eq.~\eqref{eq:C2_big}) for the Carnahan-Starling (CS) model (Eq.~\eqref{fexCS}). \textbf{(b)} Example of capacitance in the low-voltage region for the CS model. Voltage is expressed in terms of the thermal voltage $V_{T_c}= k_B\;T_c/e$ taken at $T_c$, and the capacitance is measured in units of the Helmholtz capacitance $C_H= \epsilon/4 \pi a$, where $a$ is the ion diameter. Dashed lines show the analytical approximation and the solid lines were obtained by solving Eqs.~\eqref{eq:potential}-~\eqref{eq:iondensity} numerically (see Sec.~\ref{Section:Num_Results}). \textbf{(c)}-\textbf{(d)} The same as (a)-(b) but for the lattice-gas model (Eq.~\eqref{fexBK}). In all plots, the ionophilicity $\tilde{h}_1=a^3h_1/\xi_0-\bar \rho_b=0$.}
\label{Transition_Lines}
\end{figure}

In the absence of dispersion interactions ($K=0$ in Eq.~\eqref{Eq_GP}), Eqs.~\eqref{C0} and \eqref{eq:C2_big} reduce to $ C_0= C_{D}$ and
 \begin{equation}
% \label{C2J=0}
  C_2=\frac{C_{D}}{4}\Bigg(\frac{3}{2\bar \rho_b \; \alpha(\bar{\rho}_b)}-1\Bigg),
  \label{eq:C2}
 \end{equation} 
respectively. For $K=0$, therefore, the sign of $C_2$, and thus the capacitance shape, depend only on the IL density.

For the lattice--gas model, combining Eq.~\eqref{fexBK} and Eq.~\eqref{eq:C2} gives
 \begin{equation}
  C_2=\frac{C_D}{4}\Bigg(1-3\bar \rho_b \Bigg),
 \end{equation} 
which changes sign at $\rho_b^{lg}=1/3$, implying a transformation between the bell and camel shapes at $\rho_b^{lg}$, as first pointed out by Kornyshev \cite{Kornyshev2007a}. For the CS free energy, Eq.~\eqref{fexCS}, we obtained (for $K=0$)
\begin{eqnarray}
  C_2=\frac{C_D}{4}\Bigg[\frac{3(1-\eta_b)^4}{2(1+4\eta_b+4\eta_b^2-4\eta_b^3+\eta_b^4)}-1\Bigg].
\end{eqnarray}
We have solved the equation $C_2=0$ numerically and obtained for the transition between the camel and bell shapes $\rho_b^{CS}\approx 0.098$. Note that this value is significantly lower than $\rho_b^{lg}= 1/3$ predicted by the lattice-gas model. This is similar to the critical density, which is also higher for the lattice-gas model (Fig.~\ref{Phase_Diagrams}).

Considering dispersion interactions ($K \neq 0$), Eq.~\eqref{eq:C2_big} becomes more complex and we solved it numerically using \texttt{bvp4c} routine in MATLAB$^{\textrm{\textregistered}}$~$2017a$ software. Figs.~\ref{Transition_Lines}$a,c$ show the resulting diagrams, which separate the regions of positive and negative curvatures in the low voltage capacitance. The examples of the capacitance shapes are presented in Figs.~\ref{Transition_Lines}$b,d$. This figure also demonstrates that our approximate solutions are valid only in the vicinity of $u=0$, and hence the full numerical solution is needed to describe properly the capacitance behaviour.
 
\section{Numerical results}
\label{Section:Num_Results}

We have solved Eqs.~\eqref{eq:potential} and \eqref{eq:iondensity} numerically to analyze the capacitive properties of our system in a wide range of voltages, temperatures and densities. In order to calculate the differential capacitance, we first computed the accumulated charge $Q$ according to Eq.~\eqref{Eq:Q}, and then differentiated it numerically with respect to the electrostatic potential (Eq.~\eqref{Eq:DiffCap}).

\subsection{Differential capacitance}

\begin{figure}[th]
\centering
\begin{center}
 Carnahan-Starling   
\end{center}
\vspace{-2em}
\subfloat[]{\includegraphics[scale=0.5]{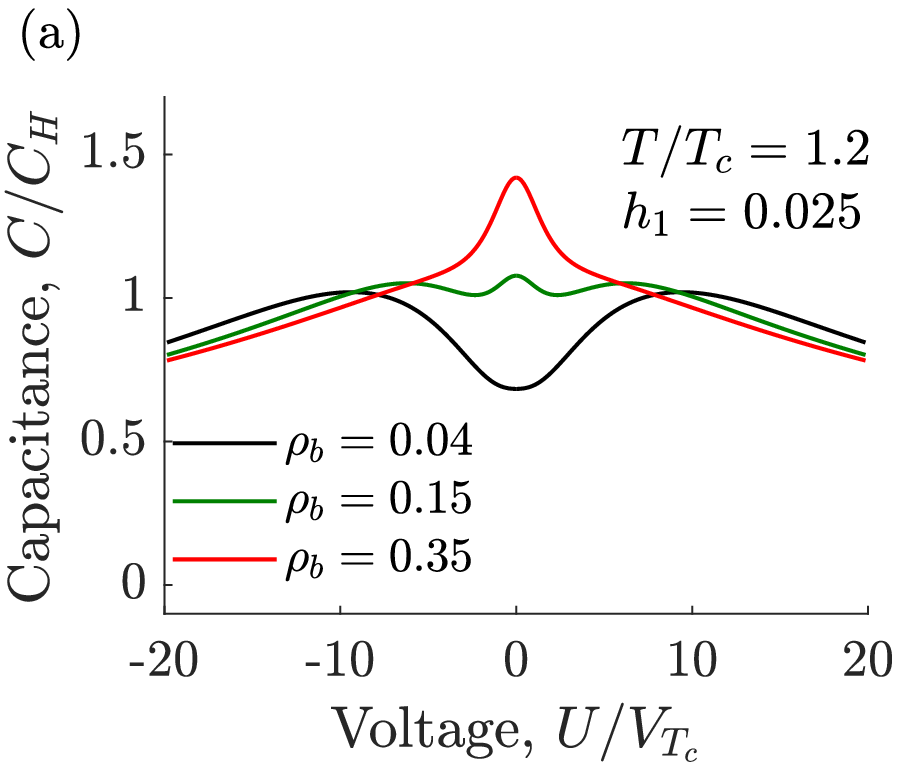}
    \label{fig:foo-1}}
    \subfloat[]{\includegraphics[scale=0.5]{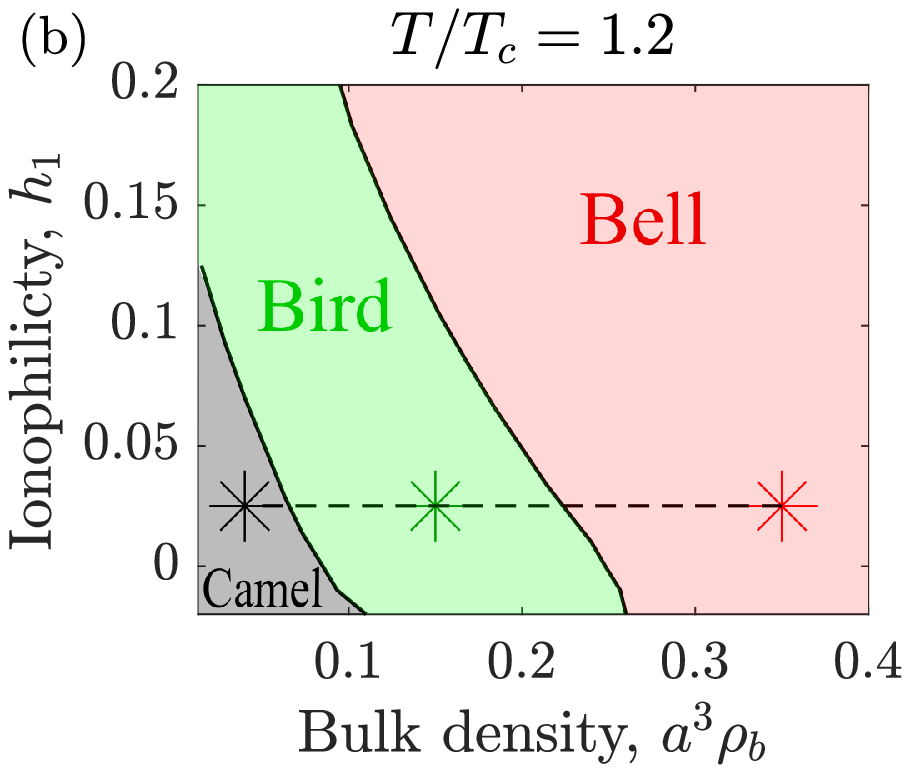}
    \label{fig:foo-1}}
\subfloat[]{\includegraphics[scale=0.5]{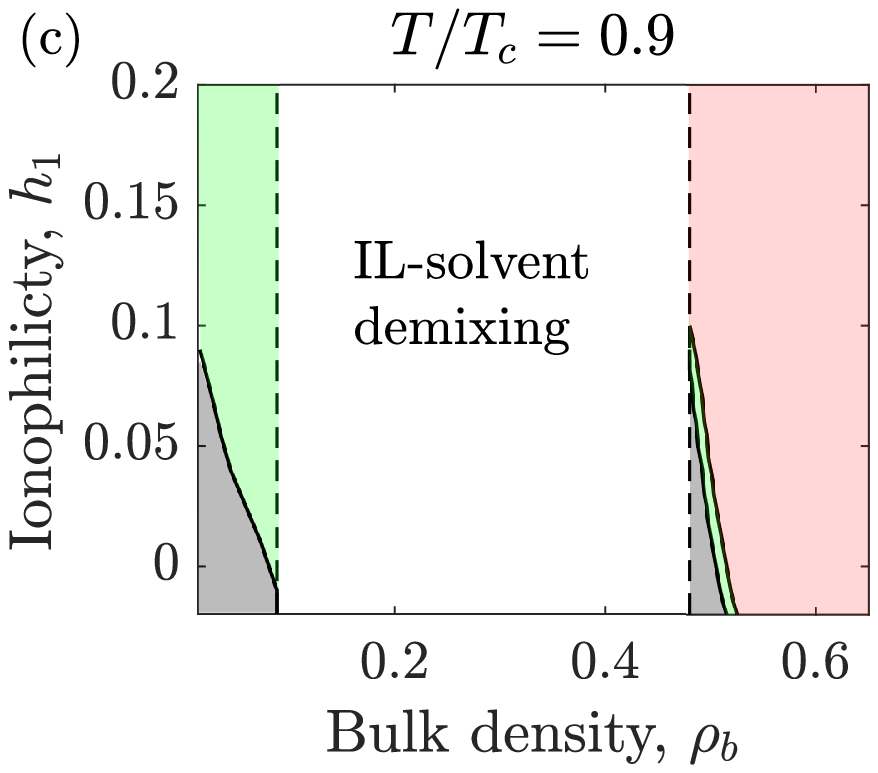}
    \label{fig:foo-1}}
\vspace{-0.1em}  
\begin{center}
Lattice-gas  
\end{center}
\vspace{-2em}
\subfloat[]{\includegraphics[scale=0.5]{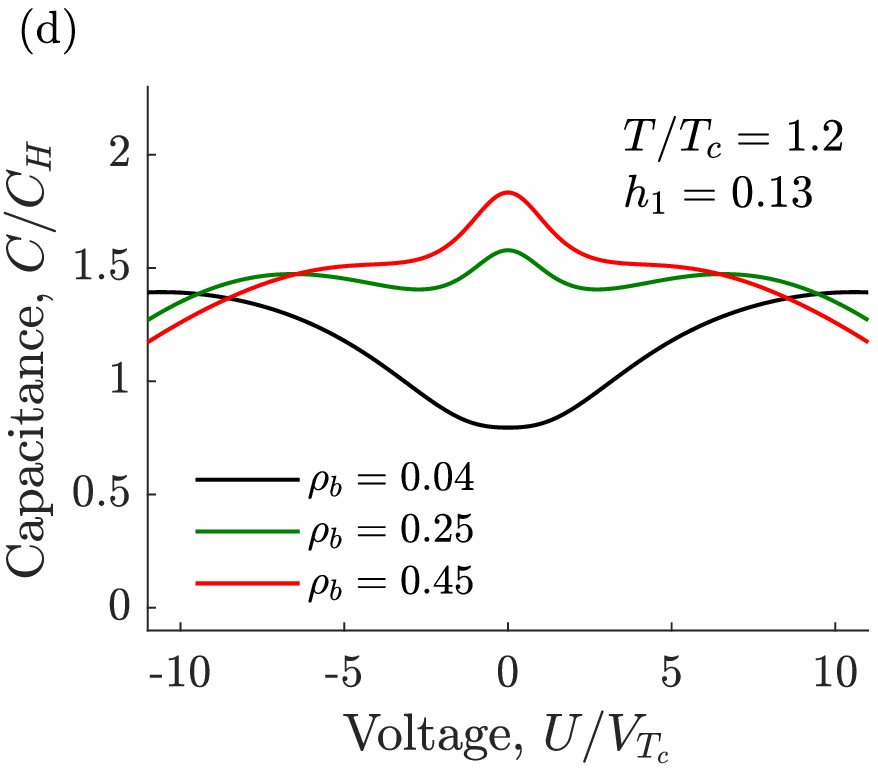}
%    \label{fig:foo-2}
}
\subfloat[]{\includegraphics[scale=0.5]{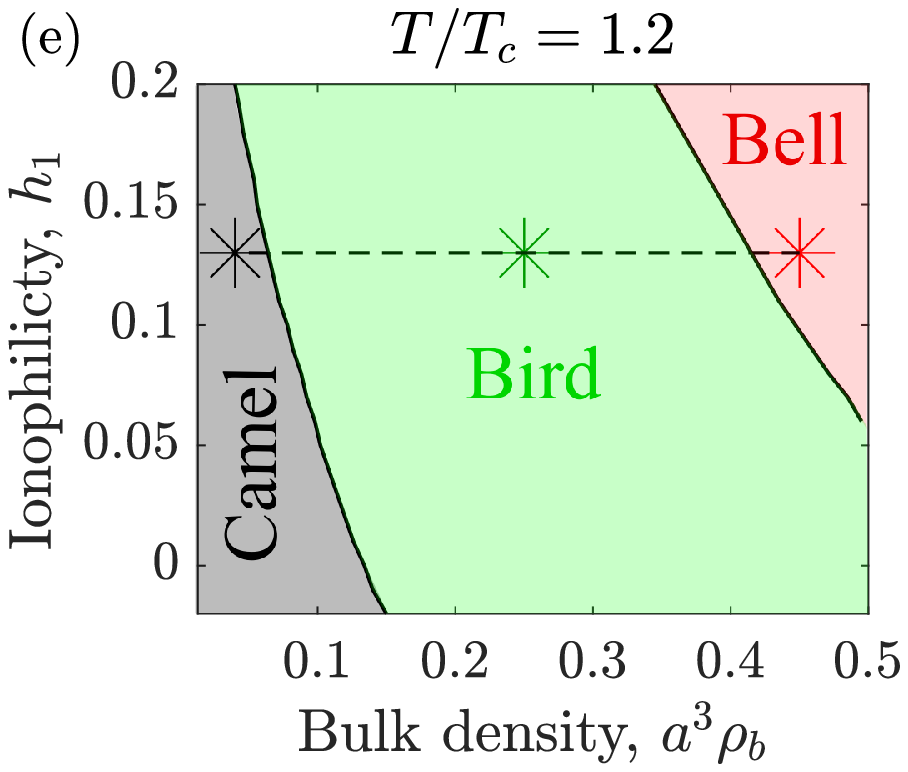}
%    \label{fig:foo-2}
}
\subfloat[]{\includegraphics[scale=0.5]{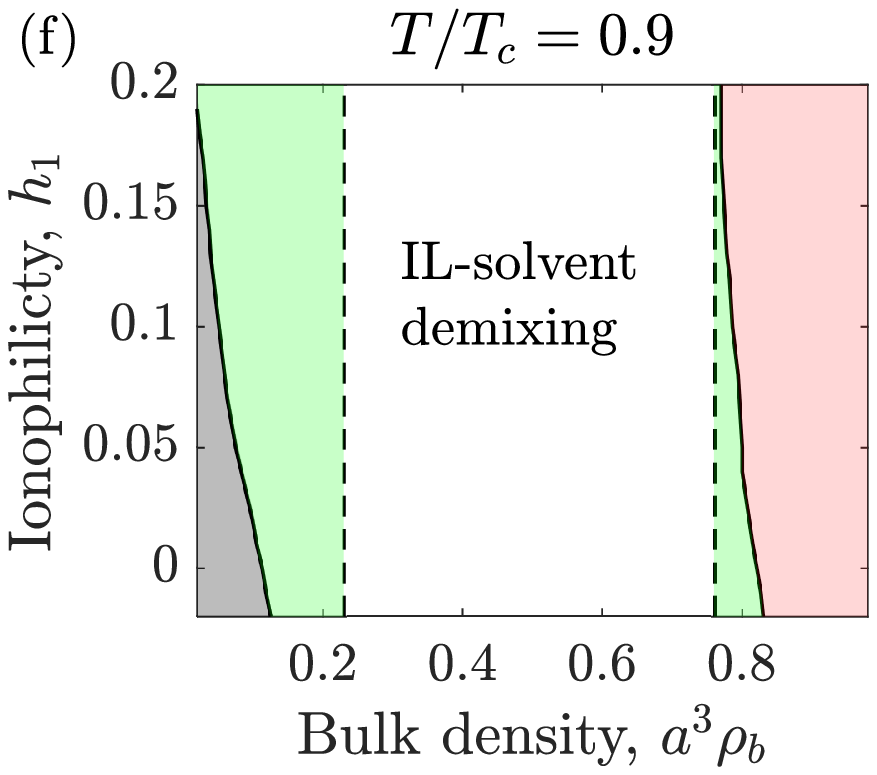}
 %   \label{fig:foo-2}
 }
\caption{\textbf{Capacitance behaviour close to demixing.} \textbf{(a)} Differential capacitance for the Carnahan-Starling (CS) entropy (Eq.~\eqref{fexCS}) as a function of applied potential for constant temperature and for a few ion concentrations, demonstrating the camel, bird, and bell-shaped capacitances. Voltage is expressed in terms of the thermal voltage $V_{T_c}= k_B\;T_c/e$ taken at $T_c$, and the capacitance is measured in units of the Helmholtz capacitance $C_H= \epsilon/4 \pi a$, where $a$ is the ion diameter \textbf{(b)}  Capacitance diagram for the CS model showing the regions of camel, bird, and bell-like capacitances at constant temperature ($T/T_c=1.2$, where $T_c$ denotes the critical point). The dashed horizontal line denotes the value of $\tilde h_1 = a^3 h_1/\xi_0$ and the symbols mark the bulk densities $\rho_b$ used in (a). \textbf{(c)}  Capacitance diagram for the CS entropy for temperature below $T_c$, $T/T_c=0.9$. The white region denotes the domain of the IL-solvent demixing (Fig.~\ref{Phase_Diagrams}). \textbf{(d)-(f)} The same as (a)-(c) but for the lattice-gas entropy.}  
\label{Capacitance_Diagram}
\end{figure}

Fig.~\ref{Capacitance_Diagram}$a$ shows that, for the CS model, there are three capacitance shapes \cite{Cruz2019a}: camel ($C_2>0$), and bird and bell shapes ($C_2<0$, compare Fig.~\ref{Transition_Lines}$a,b$). The lattice-gas model also predicts the emergence of all three capacitance shapes (Fig.~\ref{Capacitance_Diagram}$d$). This is interesting since recently \citeauthor{chen_kornyshev:18} \cite{chen_kornyshev:18} extended the steric-only lattice-gas model (\textit{i.e}, $K=0$ in Eq.~\eqref{Eq_GP}) to account for the temperature dependence and studied the capacitance in a wide range of temperature, but they did not observe the bird-like capacitance. 

Figs.~\ref{Capacitance_Diagram}$b, e$ present the capacitance diagrams for the temperatures above the critical temperature $T_c$ (\textit{i.e.}, IL--solvent is always in the mixed state). It shows the regions of the camel, bird and bell-shaped capacitances for the CS and lattice-gas models. Both models exhibit the diagrams of similar topology. For the lattice-gas model, however, the transformations between the various capacitance shapes are shifted to higher densities. This is consistent with the $K=0$ result (Fig.~\ref{Transition_Lines}), and is in similarity to the bulk phase diagram, in which the demixing region is also shifted to higher densities (Fig.~\ref{Phase_Diagrams}$b$).

Figs.~\ref{Capacitance_Diagram}$c, f$ show the capacitance diagrams for the temperature below $T_c$. Interestingly, the CS model predicts the camel shape even for high densities, but only provided the electrode is strongly ionophobic. This is likely because, close to demixing, an ionophobic electrode can induce a (macroscopically) thick layer of an ion-poor (or solvent-rich) phase, so that the system in the vicinity of the electrode behaves as being effectively dilute; note that for higher temperatures (far from demixing), only the bell shape is observed for dense ILs (Fig.~\ref{Capacitance_Diagram}$b$). It is also interesting to note that the bell and camel shapes are separated by a narrow domain of bird-like capacitance. For the lattice-gas model, however, there is no camel shape at high densities, where only the birds and bells are observed. It must be noted that our theory is inaccurate at high densities, where steric repulsions start to play the dominant role. In view of the above results, it will be interesting to study this region by more robust theories or by simulations.

\subsection{Energy storage}

\begin{figure}[t]
\centering
\includegraphics[scale=0.9]{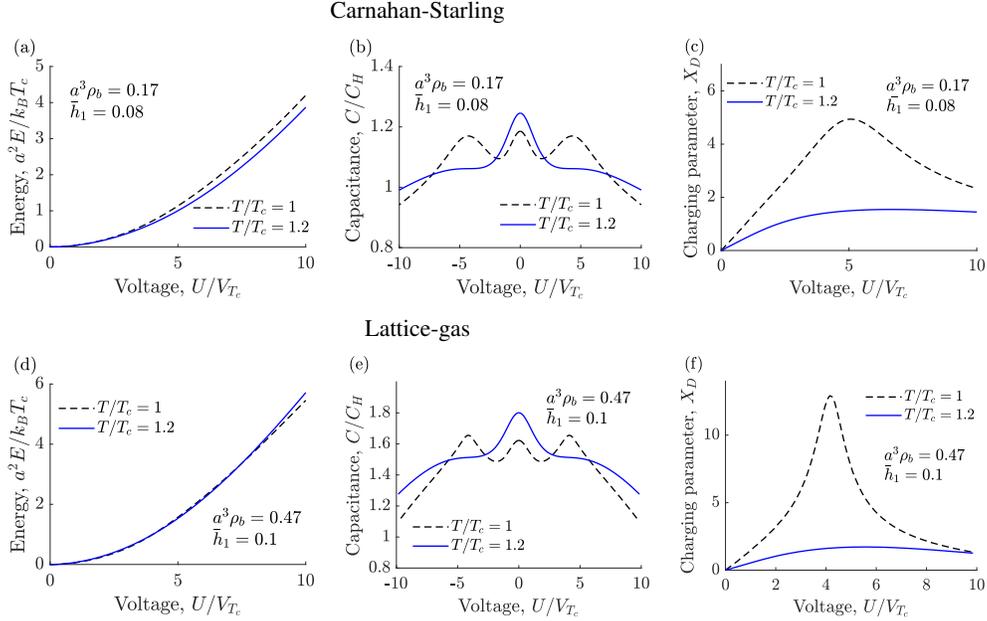}
\caption{\textbf{Energy storage and charging close to demixing.} \textbf{(a)} Stored energy per surface area for the Carnahan-Starling (CS) entropy as a function of voltage at constant ion density and electrode's ionophilicity, and for two temperatures. Energy density is measured in units of $E_T = k_BT/a^2$ taken at $T_c$, and voltage is expressed in terms of the thermal voltage $V_{T_c}= k_B\;T_c/e$ taken at $T_c$. \textbf{(b)} Differential capacitance and \textbf{(c)} charging parameter $X_D$ for the same temperatures as in (a). Capacitance is measured in units of the Helmholtz capacitance $C_H= \epsilon/4 \pi a$, where $a$ is the ion diameter. \textbf{(d)-(f)} The same as (a)-(c) but for the lattice-gas entropy.}
\label{Fig:energy}
\end{figure}

From the capacitance, we have calculated the energy density stored in an EDL as a function of the applied potential
\begin{equation}
  E(U) = \int_0^U C(u) u du . 
\end{equation}
The stored energy, together with the examples of the capacitance and charging parameter, are shown in Fig.~\ref{Fig:energy}. The charging parameter is \cite{forse:jacs:16:chmec,breitsprecher17a}
\begin{equation}
    X_D = \frac{e}{C(U)} \frac{ d\Gamma}{dU},
\end{equation}
where $\Gamma=\int_0^{\infty} \phi(z) dz$ is the surface coverage by ionic liquid.

Fig.~\ref{Fig:energy}$a$ shows the energy obtained for two temperatures for the CS entropy. In the entire range of voltages studied, the energy is higher for the lower temperature (closer to demixing). As suggested in \cite{Cruz2019a}, this temperature dependence of the stored energy can be used to generate electricity from waste heat \cite{janssen:14:prl,haertel:15:ees,wang:15:nanolett,Janssen2017,Janssen2017a, Janssen2017}. The increase of energy with decrease of temperature can be related to the capacitance behaviour. Close to demixing, capacitance increases due to the voltage-induced increase of ion density at the surface, so that the bell-shaped capacitance acquires wings and becomes bird-shaped (Fig.~\ref{Fig:energy}$b$). This is manifested in the behaviour of the charging parameter, which becomes greater than unity for the temperature close to demixing (Fig.~\ref{Fig:energy}$c$); thus, both cations and anions are adsorbed into the surface layer, leading to a strong peak in the charging parameter and capacitance. For higher voltages, however, the charging parameter decreases to $X_D \approx 1$ and the capacitance also decreases in this range of voltages.

For the lattice-gas entropy, the temperature dependence of the stored energy is weaker, as compared to the CS model (Fig.~\ref{Fig:energy}$d$). Moreover, the two curves cross each other at higher voltages, and the stored energy becomes higher for higher temperatures. Qualitatively, however, the differential capacitance (Fig.~\ref{Fig:energy}$e$) and the charging parameter (Fig.~\ref{Fig:energy}$f$) exhibit similar behaviours as for the CS model.

\section{Summary}
\label{Section:Conclusion}

We have studied electrical double layers with ionic liquid--solvent mixtures close to demixing (Fig.~\ref{Phase_Diagrams}). We proposed a model, Eq.~\eqref{Eq_GP}, appropriate for this system, and considered the Carnahan-Starling (CS) and lattice-gas expressions for the excess free energy associated with the excluded volume interactions. This model was treated both  analytically and numerically, and the results can be summarized as follows.

\begin{enumerate}

    \item Analytic expressions were obtained by using perturbation expansion, which provide good agreement with the numerical results for the charge density at low potentials (Fig.~\ref{Profiles}). 

    We also determined the capacitance shapes (at low potentials) and calculated analytically the capacitance diagrams, showing the regions of positive (camel-shaped) and negative curvatures. The transformation between these shapes can be induced by changing the ion density and temperature (Fig.~\ref{Transition_Lines}). 
    
\item \label{sum:bird_cap} Our numerical calculations revealed the emergence of a \emph{bird}-shaped capacitance, having three peaks as a function of voltage. We also calculated the capacitance diagrams, this time showing the regions with the camel, bird and bell shapes, obtained by changing the electrode's ionophilicity and bulk ion density at different fixed temperatures (Fig.~\ref{Capacitance_Diagram}). 
    
\item \label{sum:high_conc}  Interestingly, for the Carnahan-Starling free energy (here applicable for small solvent molecules), the camel-shaped capacitance, which is a signature of dilute electrolytes, can appear at high densities in the case of ionophobic electrodes; the model with the lattice-gas entropy does not exhibit the camel-like capacitance in the high density regime, however.

    \item We calculated the energy stored in an EDL at different temperatures (Fig.~\ref{Fig:energy}a,d). For the CS entropy, the energy increases when approaching demixing, which can be used to generated electricity from heat \cite{Cruz2019a,janssen:14:prl,haertel:15:ees,wang:15:nanolett,Janssen2017,Janssen2017a}. For the lattice-gas model, however, there seems to be no energy enhancement when approaching demixing. This suggests that the type of solvent (particularly the size of solvent molecules) can play an important role in the energy storage. It will be interesting to study such solvent-dependent effects in more details in future work.     
    
    \item \label{sum:bird_camel_temp} We also found that a transformation between the bell and bird shapes can be caused by varying temperature (Fig.~\ref{Fig:energy}b,e). This is due to a voltage-induced adsorption of an IL at an electrode, as manifested by the charging parameter $X_D$, which shows a strong peak at a non-zero voltage. Such an enhanced electrosorption leads to stronger screening and hence to the appearance of wings in the bell-shaped  capacitance.
    
\end{enumerate}

We presented here a simple model, and to keep it simple, we had to make a number of simplifying assumptions, in order to be able to develop some analytical insights. In particular, we assumed the cations and anions to be of the same size; we took the dielectric constant temperature and position-independent; and, most importantly, we treated the hard-core interactions on the level of the local-density approximation. While it is clear that more realistic models, or simulations, will alter the results of our calculations quantitatively, it is reasonable to expect that the qualitative behaviour \emph{is} captured by our model (particularly points \ref{sum:bird_cap} and \ref{sum:bird_camel_temp} above). Thus, our model provides the basis for further studies of electrical double layers in the vicinity of ionic liquid--solvent demixing. It will be interesting to validate our predictions by more rigorous theories, simulations and experiments, especially those obtained at high ion concentrations (e.g. point \ref{sum:high_conc} above), at which the local density approximation is known to be inaccurate \cite{antypov05a}.

Finally, we note that \citeauthor{Alam2008} \cite{Alam2008} have observed experimentally the appearance of humps at the potential of zero charge in the U-shaped capacitance for N$_2$-saturated room-temperature ILs on some electrodes; the emergence of wings in the bell-shaped capacitance was reported in a simulation study by \citeauthor{Sha2014} \cite{Sha2014} for \emph{neat} BMIM-PF$_6$ on a gold surface. Our analysis suggests that these behaviours can be related to the wetting properties of ionic liquids. It will be useful to study such relations more systematically, in order to link explicitly the wetting and electrochemical characteristics of ionic liquid-electrode systems.

\begin{acknowledgments}

	This work was supported by the European Unions Horizon 2020 research and innovation programmes under the Marie Sk\l odowska-Curie grant agreement  No.~711859 to C.C and No.~734276 to A.C. and S.K. Additional funding was received from the Ministry of Science and Higher Education of Poland for the  project No.~734276 in the years 2017-2018 (agreement No.~3854/H2020/17/2018/2) and for the implementation of the international co-financed project  No.~711859 in the  years 2017-2021. We are grateful to Yan Levin for critical reading of the manuscript and insightful comments. 

\end{acknowledgments}

\bibliography{ILsDemix}

\begin{thebibliography}{61}%
\makeatletter
\providecommand \@ifxundefined [1]{%
 \@ifx{#1\undefined}
}%
\providecommand \@ifnum [1]{%
 \ifnum #1\expandafter \@firstoftwo
 \else \expandafter \@secondoftwo
 \fi
}%
\providecommand \@ifx [1]{%
 \ifx #1\expandafter \@firstoftwo
 \else \expandafter \@secondoftwo
 \fi
}%
\providecommand \natexlab [1]{#1}%
\providecommand \enquote  [1]{``#1''}%
\providecommand \bibnamefont  [1]{#1}%
\providecommand \bibfnamefont [1]{#1}%
\providecommand \citenamefont [1]{#1}%
\providecommand \href@noop [0]{\@secondoftwo}%
\providecommand \href [0]{\begingroup \@sanitize@url \@href}%
\providecommand \@href[1]{\@@startlink{#1}\@@href}%
\providecommand \@@href[1]{\endgroup#1\@@endlink}%
\providecommand \@sanitize@url [0]{\catcode `\\12\catcode `\$12\catcode
  `\&12\catcode `\#12\catcode `\^12\catcode `\_12\catcode `\%12\relax}%
\providecommand \@@startlink[1]{}%
\providecommand \@@endlink[0]{}%
\providecommand \url  [0]{\begingroup\@sanitize@url \@url }%
\providecommand \@url [1]{\endgroup\@href {#1}{\urlprefix }}%
\providecommand \urlprefix  [0]{URL }%
\providecommand \Eprint [0]{\href }%
\providecommand \doibase [0]{http://dx.doi.org/}%
\providecommand \selectlanguage [0]{\@gobble}%
\providecommand \bibinfo  [0]{\@secondoftwo}%
\providecommand \bibfield  [0]{\@secondoftwo}%
\providecommand \translation [1]{[#1]}%
\providecommand \BibitemOpen [0]{}%
\providecommand \bibitemStop [0]{}%
\providecommand \bibitemNoStop [0]{.\EOS\space}%
\providecommand \EOS [0]{\spacefactor3000\relax}%
\providecommand \BibitemShut  [1]{\csname bibitem#1\endcsname}%
\let\auto@bib@innerbib\@empty
%</preamble>
\bibitem [{\citenamefont {Welton}(1999)}]{Welton1999}%
  \BibitemOpen
  \bibfield  {author} {\bibinfo {author} {\bibfnamefont {Thomas}\ \bibnamefont
  {Welton}},\ }\bibfield  {title} {\enquote {\bibinfo {title}
  {{Room-Temperature Ionic Liquids. Solvents for Synthesis and Catalysis}},}\
  }\href {\doibase 10.1021/cr980032t} {\bibfield  {journal} {\bibinfo
  {journal} {Chemical Reviews}\ }\textbf {\bibinfo {volume} {99}},\ \bibinfo
  {pages} {2071--2084} (\bibinfo {year} {1999})}\BibitemShut {NoStop}%
\bibitem [{\citenamefont {Fedorov}\ and\ \citenamefont
  {Kornyshev}(2014)}]{Fedorov2014}%
  \BibitemOpen
  \bibfield  {author} {\bibinfo {author} {\bibfnamefont {Maxim~V}\ \bibnamefont
  {Fedorov}}\ and\ \bibinfo {author} {\bibfnamefont {Alexei~A}\ \bibnamefont
  {Kornyshev}},\ }\bibfield  {title} {\enquote {\bibinfo {title} {{Ionic
  Liquids at Electrified Interfaces}},}\ }\href {\doibase 10.1021/cr400374x}
  {\bibfield  {journal} {\bibinfo  {journal} {Chemical Reviews}\ }\textbf
  {\bibinfo {volume} {114}},\ \bibinfo {pages} {2978--3036} (\bibinfo {year}
  {2014})}\BibitemShut {NoStop}%
\bibitem [{\citenamefont {Perkin}(2012)}]{C2CP23814D}%
  \BibitemOpen
  \bibfield  {author} {\bibinfo {author} {\bibfnamefont {Susan}\ \bibnamefont
  {Perkin}},\ }\bibfield  {title} {\enquote {\bibinfo {title} {{Ionic liquids
  in confined geometries}},}\ }\href {\doibase 10.1039/C2CP23814D} {\bibfield
  {journal} {\bibinfo  {journal} {Phys. Chem. Chem. Phys.}\ }\textbf {\bibinfo
  {volume} {14}},\ \bibinfo {pages} {5052--5062} (\bibinfo {year}
  {2012})}\BibitemShut {NoStop}%
\bibitem [{\citenamefont {Ben-Yaakov}\ \emph {et~al.}(2011)\citenamefont
  {Ben-Yaakov}, \citenamefont {Andelman}, \citenamefont {Podgornik},\ and\
  \citenamefont {Harries}}]{Ben-Yaakov2011}%
  \BibitemOpen
  \bibfield  {author} {\bibinfo {author} {\bibfnamefont {Dan}\ \bibnamefont
  {Ben-Yaakov}}, \bibinfo {author} {\bibfnamefont {David}\ \bibnamefont
  {Andelman}}, \bibinfo {author} {\bibfnamefont {Rudi}\ \bibnamefont
  {Podgornik}}, \ and\ \bibinfo {author} {\bibfnamefont {Daniel}\ \bibnamefont
  {Harries}},\ }\bibfield  {title} {\enquote {\bibinfo {title} {{Ion-specific
  hydration effects: Extending the Poisson-Boltzmann theory}},}\ }\href
  {\doibase 10.1016/J.COCIS.2011.04.012} {\bibfield  {journal} {\bibinfo
  {journal} {Current Opinion in Colloid {\&} Interface Science}\ }\textbf
  {\bibinfo {volume} {16}},\ \bibinfo {pages} {542--550} (\bibinfo {year}
  {2011})}\BibitemShut {NoStop}%
\bibitem [{\citenamefont {Parsons}(1990)}]{Parsons1990}%
  \BibitemOpen
  \bibfield  {author} {\bibinfo {author} {\bibfnamefont {Roger}\ \bibnamefont
  {Parsons}},\ }\bibfield  {title} {\enquote {\bibinfo {title} {{The electrical
  double layer: recent experimental and theoretical developments}},}\ }\href
  {\doibase 10.1021/cr00103a008} {\bibfield  {journal} {\bibinfo  {journal}
  {Chemical Reviews}\ }\textbf {\bibinfo {volume} {90}},\ \bibinfo {pages}
  {813--826} (\bibinfo {year} {1990})}\BibitemShut {NoStop}%
\bibitem [{\citenamefont {Bikerman}(1942)}]{Bikerman1942}%
  \BibitemOpen
  \bibfield  {author} {\bibinfo {author} {\bibfnamefont {J.J.}\ \bibnamefont
  {Bikerman}},\ }\bibfield  {title} {\enquote {\bibinfo {title} {{XXXIX.
  Structure and capacity of electrical double layer}},}\ }\href {\doibase
  10.1080/14786444208520813} {\bibfield  {journal} {\bibinfo  {journal} {The
  London, Edinburgh, and Dublin Philosophical Magazine and Journal of Science}\
  }\textbf {\bibinfo {volume} {33}},\ \bibinfo {pages} {384--397} (\bibinfo
  {year} {1942})}\BibitemShut {NoStop}%
\bibitem [{\citenamefont {Wicke}\ and\ \citenamefont
  {Eigen}(1952)}]{wicke1952einfluss}%
  \BibitemOpen
  \bibfield  {author} {\bibinfo {author} {\bibfnamefont {E}~\bibnamefont
  {Wicke}}\ and\ \bibinfo {author} {\bibfnamefont {M}~\bibnamefont {Eigen}},\
  }\bibfield  {title} {\enquote {\bibinfo {title} {{\"U}ber den einflu{\ss} des
  raumbedarfs von ionen in w{\"a}{\ss}riger l{\"o}sung auf ihre verteilung in
  elektrischen feld und ihre aktivit{\"a}tskoeffizienten},}\ }\href@noop {}
  {\bibfield  {journal} {\bibinfo  {journal} {Zeitschrift f{\"u}r
  Elektrochemie, Berichte der Bunsengesellschaft f{\"u}r physikalische Chemie}\
  }\textbf {\bibinfo {volume} {56}},\ \bibinfo {pages} {551--561} (\bibinfo
  {year} {1952})}\BibitemShut {NoStop}%
\bibitem [{\citenamefont {Freise}(1952)}]{Freise1952}%
  \BibitemOpen
  \bibfield  {author} {\bibinfo {author} {\bibfnamefont {V.}~\bibnamefont
  {Freise}},\ }\bibfield  {title} {\enquote {\bibinfo {title} {{Zur Theorie der
  diffusen Doppelschicht}},}\ }\href@noop {} {\bibfield  {journal} {\bibinfo
  {journal} {Zeitschrift f{\"{u}}r Elektrochemie}\ }\textbf {\bibinfo {volume}
  {56}},\ \bibinfo {pages} {822--827} (\bibinfo {year} {1952})}\BibitemShut
  {NoStop}%
\bibitem [{\citenamefont {Eigen}\ and\ \citenamefont
  {Wicke}(1954)}]{Eigen1954}%
  \BibitemOpen
  \bibfield  {author} {\bibinfo {author} {\bibfnamefont {M.}~\bibnamefont
  {Eigen}}\ and\ \bibinfo {author} {\bibfnamefont {E.}~\bibnamefont {Wicke}},\
  }\bibfield  {title} {\enquote {\bibinfo {title} {{The thermodynamics of
  electrolytes at higher concentration}},}\ }\href {\doibase
  10.1021/j150519a007} {\bibfield  {journal} {\bibinfo  {journal} {J. Phys.
  Chem.}\ }\textbf {\bibinfo {volume} {58}},\ \bibinfo {pages} {702--714}
  (\bibinfo {year} {1954})}\BibitemShut {NoStop}%
\bibitem [{\citenamefont {Borukhov}\ \emph {et~al.}(1997)\citenamefont
  {Borukhov}, \citenamefont {Andelman},\ and\ \citenamefont
  {Orland}}]{PhysRevLett.79.435}%
  \BibitemOpen
  \bibfield  {author} {\bibinfo {author} {\bibfnamefont {Itamar}\ \bibnamefont
  {Borukhov}}, \bibinfo {author} {\bibfnamefont {David}\ \bibnamefont
  {Andelman}}, \ and\ \bibinfo {author} {\bibfnamefont {Henri}\ \bibnamefont
  {Orland}},\ }\bibfield  {title} {\enquote {\bibinfo {title} {{Steric Effects
  in Electrolytes: A Modified Poisson-Boltzmann Equation}},}\ }\href {\doibase
  10.1103/PhysRevLett.79.435} {\bibfield  {journal} {\bibinfo  {journal} {Phys.
  Rev. Lett.}\ }\textbf {\bibinfo {volume} {79}},\ \bibinfo {pages} {435--438}
  (\bibinfo {year} {1997})}\BibitemShut {NoStop}%
\bibitem [{\citenamefont {di~Caprio}\ \emph {et~al.}(2003)\citenamefont
  {di~Caprio}, \citenamefont {Borkowska},\ and\ \citenamefont
  {Stafiej}}]{DiCaprio2003}%
  \BibitemOpen
  \bibfield  {author} {\bibinfo {author} {\bibfnamefont {D.}~\bibnamefont
  {di~Caprio}}, \bibinfo {author} {\bibfnamefont {Z.}~\bibnamefont
  {Borkowska}}, \ and\ \bibinfo {author} {\bibfnamefont {J.}~\bibnamefont
  {Stafiej}},\ }\bibfield  {title} {\enquote {\bibinfo {title} {{Simple
  extension of the Gouy–Chapman theory including hard sphere effects.}}}\
  }\href {\doibase 10.1016/s0022-0728(02)01270-6} {\bibfield  {journal}
  {\bibinfo  {journal} {Journal of Electroanalytical Chemistry}\ }\textbf
  {\bibinfo {volume} {540}},\ \bibinfo {pages} {17--23} (\bibinfo {year}
  {2003})}\BibitemShut {NoStop}%
\bibitem [{\citenamefont {Antypov}\ \emph {et~al.}(2005)\citenamefont
  {Antypov}, \citenamefont {Barbosa},\ and\ \citenamefont {Holm}}]{antypov05a}%
  \BibitemOpen
  \bibfield  {author} {\bibinfo {author} {\bibfnamefont {D.}~\bibnamefont
  {Antypov}}, \bibinfo {author} {\bibfnamefont {M.~C.}\ \bibnamefont
  {Barbosa}}, \ and\ \bibinfo {author} {\bibfnamefont {C.}~\bibnamefont
  {Holm}},\ }\bibfield  {title} {\enquote {\bibinfo {title} {A simple non-local
  approach to treat size correlations within {Poisson-Boltzmann} theory},}\
  }\href@noop {} {\bibfield  {journal} {\bibinfo  {journal} {Phys. Rev. E}\
  }\textbf {\bibinfo {volume} {71}},\ \bibinfo {pages} {061106} (\bibinfo
  {year} {2005})}\BibitemShut {NoStop}%
\bibitem [{\citenamefont {Oldham}(2008)}]{Oldham2008}%
  \BibitemOpen
  \bibfield  {author} {\bibinfo {author} {\bibfnamefont {Keith~B.}\
  \bibnamefont {Oldham}},\ }\bibfield  {title} {\enquote {\bibinfo {title} {{A
  Gouy-Chapman-Stern model of the double layer at a (metal)/(ionic liquid)
  interface}},}\ }\href {\doibase 10.1016/j.jelechem.2007.10.017} {\bibfield
  {journal} {\bibinfo  {journal} {Journal of Electroanalytical Chemistry}\
  }\textbf {\bibinfo {volume} {613}},\ \bibinfo {pages} {131--138} (\bibinfo
  {year} {2008})}\BibitemShut {NoStop}%
\bibitem [{\citenamefont {McEldrew}\ \emph {et~al.}(2018)\citenamefont
  {McEldrew}, \citenamefont {Goodwin}, \citenamefont {Kornyshev},\ and\
  \citenamefont {Bazant}}]{McEldrew2018}%
  \BibitemOpen
  \bibfield  {author} {\bibinfo {author} {\bibfnamefont {Michael}\ \bibnamefont
  {McEldrew}}, \bibinfo {author} {\bibfnamefont {Zachary~A.H.}\ \bibnamefont
  {Goodwin}}, \bibinfo {author} {\bibfnamefont {Alexei~A.}\ \bibnamefont
  {Kornyshev}}, \ and\ \bibinfo {author} {\bibfnamefont {Martin~Z.}\
  \bibnamefont {Bazant}},\ }\bibfield  {title} {\enquote {\bibinfo {title}
  {{Theory of the Double Layer in Water-in-Salt Electrolytes}},}\ }\href
  {\doibase 10.1021/acs.jpclett.8b02543} {\bibfield  {journal} {\bibinfo
  {journal} {Journal of Physical Chemistry Letters}\ }\textbf {\bibinfo
  {volume} {9}},\ \bibinfo {pages} {5840--5846} (\bibinfo {year}
  {2018})}\BibitemShut {NoStop}%
\bibitem [{\citenamefont {Bohinc}\ \emph {et~al.}(2001)\citenamefont {Bohinc},
  \citenamefont {Kralj-Igli{\v{c}}},\ and\ \citenamefont
  {Igli{\v{c}}}}]{Bohinc2001a}%
  \BibitemOpen
  \bibfield  {author} {\bibinfo {author} {\bibfnamefont {Klemen}\ \bibnamefont
  {Bohinc}}, \bibinfo {author} {\bibfnamefont {Veronika}\ \bibnamefont
  {Kralj-Igli{\v{c}}}}, \ and\ \bibinfo {author} {\bibfnamefont {Aleš}\
  \bibnamefont {Igli{\v{c}}}},\ }\bibfield  {title} {\enquote {\bibinfo {title}
  {{Thickness of electrical double layer. Effect of ion size}},}\ }\href
  {\doibase 10.1016/S0013-4686(01)00525-4} {\bibfield  {journal} {\bibinfo
  {journal} {Electrochimica Acta}\ }\textbf {\bibinfo {volume} {46}},\ \bibinfo
  {pages} {3033--3040} (\bibinfo {year} {2001})}\BibitemShut {NoStop}%
\bibitem [{\citenamefont {Kornyshev}(2007)}]{Kornyshev2007a}%
  \BibitemOpen
  \bibfield  {author} {\bibinfo {author} {\bibfnamefont {Alexei~A}\
  \bibnamefont {Kornyshev}},\ }\bibfield  {title} {\enquote {\bibinfo {title}
  {{Double-Layer in Ionic Liquids: Paradigm Change?}}}\ }\href {\doibase
  10.1021/jp067857o} {\bibfield  {journal} {\bibinfo  {journal} {The Journal of
  Physical Chemistry B}\ }\textbf {\bibinfo {volume} {111}},\ \bibinfo {pages}
  {5545--5557} (\bibinfo {year} {2007})}\BibitemShut {NoStop}%
\bibitem [{\citenamefont {Kilic}\ \emph {et~al.}(2007)\citenamefont {Kilic},
  \citenamefont {Bazant},\ and\ \citenamefont {Ajdari}}]{kilic:pre:07a}%
  \BibitemOpen
  \bibfield  {author} {\bibinfo {author} {\bibfnamefont {Mustafa~Sabri}\
  \bibnamefont {Kilic}}, \bibinfo {author} {\bibfnamefont {Martin~Z.}\
  \bibnamefont {Bazant}}, \ and\ \bibinfo {author} {\bibfnamefont {Armand}\
  \bibnamefont {Ajdari}},\ }\bibfield  {title} {\enquote {\bibinfo {title}
  {Steric effects in the dynamics of electrolytes at large applied voltages. i.
  double-layer charging},}\ }\href@noop {} {\bibfield  {journal} {\bibinfo
  {journal} {Phys. Rev. E}\ }\textbf {\bibinfo {volume} {75}},\ \bibinfo
  {pages} {021502} (\bibinfo {year} {2007})}\BibitemShut {NoStop}%
\bibitem [{\citenamefont {Frydel}\ and\ \citenamefont
  {Levin}(2012)}]{Frydel2012}%
  \BibitemOpen
  \bibfield  {author} {\bibinfo {author} {\bibfnamefont {Derek}\ \bibnamefont
  {Frydel}}\ and\ \bibinfo {author} {\bibfnamefont {Yan}\ \bibnamefont
  {Levin}},\ }\bibfield  {title} {\enquote {\bibinfo {title} {A close look into
  the excluded volume effects within a double layer},}\ }\href {\doibase
  10.1063/1.4761938} {\bibfield  {journal} {\bibinfo  {journal} {J. Chem.
  Phys.}\ }\textbf {\bibinfo {volume} {137}},\ \bibinfo {pages} {164703}
  (\bibinfo {year} {2012})}\BibitemShut {NoStop}%
\bibitem [{\citenamefont {Minton}\ and\ \citenamefont
  {Lue}(2016)}]{Minton2016}%
  \BibitemOpen
  \bibfield  {author} {\bibinfo {author} {\bibfnamefont {Geraint}\ \bibnamefont
  {Minton}}\ and\ \bibinfo {author} {\bibfnamefont {Leo}\ \bibnamefont {Lue}},\
  }\bibfield  {title} {\enquote {\bibinfo {title} {{The influence of excluded
  volume and excess ion polarisability on the capacitance of the electric
  double layer}},}\ }\href {\doibase 10.1080/00268976.2016.1169327} {\bibfield
  {journal} {\bibinfo  {journal} {Molecular Physics}\ }\textbf {\bibinfo
  {volume} {114}},\ \bibinfo {pages} {2477--2491} (\bibinfo {year}
  {2016})}\BibitemShut {NoStop}%
\bibitem [{\citenamefont {Girotto}\ \emph {et~al.}(2018)\citenamefont
  {Girotto}, \citenamefont {Malossi}, \citenamefont {dos Santos},\ and\
  \citenamefont {Levin}}]{Girotto2018}%
  \BibitemOpen
  \bibfield  {author} {\bibinfo {author} {\bibfnamefont {Matheus}\ \bibnamefont
  {Girotto}}, \bibinfo {author} {\bibfnamefont {Rodrigo~M.}\ \bibnamefont
  {Malossi}}, \bibinfo {author} {\bibfnamefont {Alexandre~P.}\ \bibnamefont
  {dos Santos}}, \ and\ \bibinfo {author} {\bibfnamefont {Yan}\ \bibnamefont
  {Levin}},\ }\bibfield  {title} {\enquote {\bibinfo {title} {Lattice model of
  ionic liquid confined by metal electrodes},}\ }\href {\doibase
  10.1063/1.5013337} {\bibfield  {journal} {\bibinfo  {journal} {J. Chem.
  Phys.}\ }\textbf {\bibinfo {volume} {148}},\ \bibinfo {pages} {193829}
  (\bibinfo {year} {2018})}\BibitemShut {NoStop}%
\bibitem [{\citenamefont {Lockett}\ \emph
  {et~al.}(2008{\natexlab{a}})\citenamefont {Lockett}, \citenamefont {Sedev},
  \citenamefont {Ralston}, \citenamefont {Horne},\ and\ \citenamefont
  {Rodopoulos}}]{Lockett2008b}%
  \BibitemOpen
  \bibfield  {author} {\bibinfo {author} {\bibfnamefont {Vera}\ \bibnamefont
  {Lockett}}, \bibinfo {author} {\bibfnamefont {Rossen}\ \bibnamefont {Sedev}},
  \bibinfo {author} {\bibfnamefont {John}\ \bibnamefont {Ralston}}, \bibinfo
  {author} {\bibfnamefont {Mike}\ \bibnamefont {Horne}}, \ and\ \bibinfo
  {author} {\bibfnamefont {Theo}\ \bibnamefont {Rodopoulos}},\ }\bibfield
  {title} {\enquote {\bibinfo {title} {{Differential Capacitance of the
  Electrical Double Layer in Imidazolium-Based Ionic Liquids: Influence of
  Potential, Cation Size, and Temperature}},}\ }\href {\doibase
  10.1021/jp7100732} {\bibfield  {journal} {\bibinfo  {journal} {The Journal of
  Physical Chemistry C}\ }\textbf {\bibinfo {volume} {112}},\ \bibinfo {pages}
  {7486--7495} (\bibinfo {year} {2008}{\natexlab{a}})}\BibitemShut {NoStop}%
\bibitem [{\citenamefont {Silva}\ \emph {et~al.}(2008)\citenamefont {Silva},
  \citenamefont {Gomes}, \citenamefont {Figueiredo}, \citenamefont {Costa},
  \citenamefont {Martins},\ and\ \citenamefont {Pereira}}]{Silva2008}%
  \BibitemOpen
  \bibfield  {author} {\bibinfo {author} {\bibfnamefont {Fernando}\
  \bibnamefont {Silva}}, \bibinfo {author} {\bibfnamefont {Cristiana}\
  \bibnamefont {Gomes}}, \bibinfo {author} {\bibfnamefont {Marta}\ \bibnamefont
  {Figueiredo}}, \bibinfo {author} {\bibfnamefont {Renata}\ \bibnamefont
  {Costa}}, \bibinfo {author} {\bibfnamefont {Ana}\ \bibnamefont {Martins}}, \
  and\ \bibinfo {author} {\bibfnamefont {Carlos~M.}\ \bibnamefont {Pereira}},\
  }\bibfield  {title} {\enquote {\bibinfo {title} {{The electrical double layer
  at the [BMIM][PF6] ionic liquid/electrode interface - Effect of temperature
  on the differential capacitance}},}\ }\href {\doibase
  10.1016/j.jelechem.2008.05.014} {\bibfield  {journal} {\bibinfo  {journal}
  {Journal of Electroanalytical Chemistry}\ }\textbf {\bibinfo {volume}
  {622}},\ \bibinfo {pages} {153--160} (\bibinfo {year} {2008})}\BibitemShut
  {NoStop}%
\bibitem [{\citenamefont {Dr{\"{u}}schler}\ \emph {et~al.}(2012)\citenamefont
  {Dr{\"{u}}schler}, \citenamefont {Borisenko}, \citenamefont {Wallauer},
  \citenamefont {Winter}, \citenamefont {Huber}, \citenamefont {Endres},\ and\
  \citenamefont {Roling}}]{Druschler2012a}%
  \BibitemOpen
  \bibfield  {author} {\bibinfo {author} {\bibfnamefont {Marcel}\ \bibnamefont
  {Dr{\"{u}}schler}}, \bibinfo {author} {\bibfnamefont {Natalia}\ \bibnamefont
  {Borisenko}}, \bibinfo {author} {\bibfnamefont {Jens}\ \bibnamefont
  {Wallauer}}, \bibinfo {author} {\bibfnamefont {Christian}\ \bibnamefont
  {Winter}}, \bibinfo {author} {\bibfnamefont {Benedikt}\ \bibnamefont
  {Huber}}, \bibinfo {author} {\bibfnamefont {Frank}\ \bibnamefont {Endres}}, \
  and\ \bibinfo {author} {\bibfnamefont {Bernhard}\ \bibnamefont {Roling}},\
  }\bibfield  {title} {\enquote {\bibinfo {title} {{New insights into the
  interface between a single-crystalline metal electrode and an extremely pure
  ionic liquid: slow interfacial processes and the influence of temperature on
  interfacial dynamics}},}\ }\href {\doibase 10.1039/C2CP40288B} {\bibfield
  {journal} {\bibinfo  {journal} {Physical Chemistry Chemical Physics}\
  }\textbf {\bibinfo {volume} {14}},\ \bibinfo {pages} {5090--5099} (\bibinfo
  {year} {2012})}\BibitemShut {NoStop}%
\bibitem [{\citenamefont {Ivani{\v{s}}t{\v{s}}ev}\ \emph
  {et~al.}(2017)\citenamefont {Ivani{\v{s}}t{\v{s}}ev}, \citenamefont
  {Kirchner},\ and\ \citenamefont {Fedorov}}]{Ivanistsev2017a}%
  \BibitemOpen
  \bibfield  {author} {\bibinfo {author} {\bibfnamefont {Vladislav~B}\
  \bibnamefont {Ivani{\v{s}}t{\v{s}}ev}}, \bibinfo {author} {\bibfnamefont
  {Kathleen}\ \bibnamefont {Kirchner}}, \ and\ \bibinfo {author} {\bibfnamefont
  {Maxim~V.}\ \bibnamefont {Fedorov}},\ }\bibfield  {title} {\enquote {\bibinfo
  {title} {{Double layer in ionic liquids: capacitance vs temperature}},}\
  }\href {http://arxiv.org/abs/1711.06854} {\  (\bibinfo {year}
  {2017})}\BibitemShut {NoStop}%
\bibitem [{\citenamefont {Holovko}\ \emph {et~al.}(2001)\citenamefont
  {Holovko}, \citenamefont {Kapko}, \citenamefont {Henderson},\ and\
  \citenamefont {Boda}}]{holovko:01}%
  \BibitemOpen
  \bibfield  {author} {\bibinfo {author} {\bibfnamefont {Myroslav}\
  \bibnamefont {Holovko}}, \bibinfo {author} {\bibfnamefont {Vitalyj}\
  \bibnamefont {Kapko}}, \bibinfo {author} {\bibfnamefont {Douglas}\
  \bibnamefont {Henderson}}, \ and\ \bibinfo {author} {\bibfnamefont {Dezsö}\
  \bibnamefont {Boda}},\ }\bibfield  {title} {\enquote {\bibinfo {title} {{On
  the Influence of Ionic Association on the Capacitance of an Electrical Double
  Layer}},}\ }\href {\doibase 10.1016/S0009-2614(01)00505-X} {\bibfield
  {journal} {\bibinfo  {journal} {Chem. Phys. Lett.}\ }\textbf {\bibinfo
  {volume} {341}},\ \bibinfo {pages} {363--368} (\bibinfo {year}
  {2001})}\BibitemShut {NoStop}%
\bibitem [{\citenamefont {Reszko-Zygmunt}\ \emph {et~al.}(2005)\citenamefont
  {Reszko-Zygmunt}, \citenamefont {Soko{\l}owski}, \citenamefont {Henderson},\
  and\ \citenamefont {Boda}}]{Reszko-Zygmunt2005}%
  \BibitemOpen
  \bibfield  {author} {\bibinfo {author} {\bibfnamefont {J.}~\bibnamefont
  {Reszko-Zygmunt}}, \bibinfo {author} {\bibfnamefont {S.}~\bibnamefont
  {Soko{\l}owski}}, \bibinfo {author} {\bibfnamefont {D.}~\bibnamefont
  {Henderson}}, \ and\ \bibinfo {author} {\bibfnamefont {D.}~\bibnamefont
  {Boda}},\ }\bibfield  {title} {\enquote {\bibinfo {title} {{Temperature
  dependence of the double layer capacitance for the restricted primitive model
  of an electrolyte solution from a density functional approach}},}\ }\href
  {\doibase 10.1063/1.1850453} {\bibfield  {journal} {\bibinfo  {journal}
  {Journal of Chemical Physics}\ }\textbf {\bibinfo {volume} {122}},\ \bibinfo
  {pages} {084504} (\bibinfo {year} {2005})}\BibitemShut {NoStop}%
\bibitem [{\citenamefont {Chen}\ \emph {et~al.}(2018)\citenamefont {Chen},
  \citenamefont {Goodwin}, \citenamefont {Feng},\ and\ \citenamefont
  {Kornyshev}}]{chen_kornyshev:18}%
  \BibitemOpen
  \bibfield  {author} {\bibinfo {author} {\bibfnamefont {Ming}\ \bibnamefont
  {Chen}}, \bibinfo {author} {\bibfnamefont {Zachary A~H}\ \bibnamefont
  {Goodwin}}, \bibinfo {author} {\bibfnamefont {Guang}\ \bibnamefont {Feng}}, \
  and\ \bibinfo {author} {\bibfnamefont {Alexei~A}\ \bibnamefont {Kornyshev}},\
  }\bibfield  {title} {\enquote {\bibinfo {title} {{On the Temperature
  Dependence of the Double Layer Capacitance of Ionic Liquids}},}\ }\href
  {\doibase 10.1016/j.jelechem.2017.11.005} {\bibfield  {journal} {\bibinfo
  {journal} {J. Electroanal. Chem.}\ }\textbf {\bibinfo {volume} {819}},\
  \bibinfo {pages} {347--358} (\bibinfo {year} {2018})}\BibitemShut {NoStop}%
\bibitem [{\citenamefont {Vatamanu}\ \emph {et~al.}(2011)\citenamefont
  {Vatamanu}, \citenamefont {Borodin},\ and\ \citenamefont
  {Smith}}]{Vatamanu2011a}%
  \BibitemOpen
  \bibfield  {author} {\bibinfo {author} {\bibfnamefont {Jenel}\ \bibnamefont
  {Vatamanu}}, \bibinfo {author} {\bibfnamefont {Oleg}\ \bibnamefont
  {Borodin}}, \ and\ \bibinfo {author} {\bibfnamefont {Grant~D}\ \bibnamefont
  {Smith}},\ }\bibfield  {title} {\enquote {\bibinfo {title} {{Molecular
  Simulations of the Electric Double Layer Structure, Differential Capacitance,
  and Charging Kinetics for N-Methyl-N-propylpyrrolidinium
  Bis(fluorosulfonyl)imide at Graphite Electrodes}},}\ }\href {\doibase
  10.1021/jp2001207} {\bibfield  {journal} {\bibinfo  {journal} {The Journal of
  Physical Chemistry B}\ }\textbf {\bibinfo {volume} {115}},\ \bibinfo {pages}
  {3073--3084} (\bibinfo {year} {2011})}\BibitemShut {NoStop}%
\bibitem [{\citenamefont {Bonn}\ \emph {et~al.}(2009)\citenamefont {Bonn},
  \citenamefont {Eggers}, \citenamefont {Indekeu}, \citenamefont {Meunier},\
  and\ \citenamefont {Rolley}}]{Bonn2009}%
  \BibitemOpen
  \bibfield  {author} {\bibinfo {author} {\bibfnamefont {Daniel}\ \bibnamefont
  {Bonn}}, \bibinfo {author} {\bibfnamefont {Jens}\ \bibnamefont {Eggers}},
  \bibinfo {author} {\bibfnamefont {Joseph}\ \bibnamefont {Indekeu}}, \bibinfo
  {author} {\bibfnamefont {Jacques}\ \bibnamefont {Meunier}}, \ and\ \bibinfo
  {author} {\bibfnamefont {Etienne}\ \bibnamefont {Rolley}},\ }\bibfield
  {title} {\enquote {\bibinfo {title} {{Wetting and spreading}},}\ }\href
  {\doibase 10.1103/revmodphys.81.739} {\bibfield  {journal} {\bibinfo
  {journal} {Rev. Mod. Phys.}\ }\textbf {\bibinfo {volume} {81}},\ \bibinfo
  {pages} {739--805} (\bibinfo {year} {2009})}\BibitemShut {NoStop}%
\bibitem [{\citenamefont {Cruz}\ \emph {et~al.}(2019)\citenamefont {Cruz},
  \citenamefont {Ciach}, \citenamefont {Lomba},\ and\ \citenamefont
  {Kondrat}}]{Cruz2019a}%
  \BibitemOpen
  \bibfield  {author} {\bibinfo {author} {\bibfnamefont {Carolina}\
  \bibnamefont {Cruz}}, \bibinfo {author} {\bibfnamefont {Alina}\ \bibnamefont
  {Ciach}}, \bibinfo {author} {\bibfnamefont {Enrique}\ \bibnamefont {Lomba}},
  \ and\ \bibinfo {author} {\bibfnamefont {Svyatoslav}\ \bibnamefont
  {Kondrat}},\ }\bibfield  {title} {\enquote {\bibinfo {title} {{Electrical
  Double Layers Close to Ionic Liquid-Solvent Demixing}},}\ }\href {\doibase
  10.1021/acs.jpcc.8b09772} {\bibfield  {journal} {\bibinfo  {journal} {Journal
  of Physical Chemistry C}\ }\textbf {\bibinfo {volume} {123}},\ \bibinfo
  {pages} {1596--1601} (\bibinfo {year} {2019})}\BibitemShut {NoStop}%
\bibitem [{\citenamefont {Elshwishin}\ \emph {et~al.}(2014)\citenamefont
  {Elshwishin}, \citenamefont {K{\"{o}}ser}, \citenamefont {Schr{\"{o}}er},\
  and\ \citenamefont {Qiao}}]{Elshwishin2014}%
  \BibitemOpen
  \bibfield  {author} {\bibinfo {author} {\bibfnamefont {A.}~\bibnamefont
  {Elshwishin}}, \bibinfo {author} {\bibfnamefont {J.}~\bibnamefont
  {K{\"{o}}ser}}, \bibinfo {author} {\bibfnamefont {W.}~\bibnamefont
  {Schr{\"{o}}er}}, \ and\ \bibinfo {author} {\bibfnamefont {Baofu}\
  \bibnamefont {Qiao}},\ }\bibfield  {title} {\enquote {\bibinfo {title}
  {{Liquid-liquid phase separation of ionic liquids in solutions: Ionic liquids
  with the triflat anion solved in aryl halides}},}\ }\href {\doibase
  10.1016/j.molliq.2013.07.012} {\bibfield  {journal} {\bibinfo  {journal}
  {Journal of Molecular Liquids}\ }\textbf {\bibinfo {volume} {192}},\ \bibinfo
  {pages} {127--136} (\bibinfo {year} {2014})}\BibitemShut {NoStop}%
\bibitem [{\citenamefont {Kralj-Igli{\v{c}}}\ and\ \citenamefont
  {Igli{\v{c}}}(1996)}]{Kralj-Igli1996}%
  \BibitemOpen
  \bibfield  {author} {\bibinfo {author} {\bibfnamefont {Veronika}\
  \bibnamefont {Kralj-Igli{\v{c}}}}\ and\ \bibinfo {author} {\bibfnamefont
  {Ale{\v{s}}}\ \bibnamefont {Igli{\v{c}}}},\ }\bibfield  {title} {\enquote
  {\bibinfo {title} {{A Simple Statistical Mechanical Approach to the free
  Energy of the Electric Double Layer Including the Excluded Volume Effect}},}\
  }\href {\doibase 10.1051/jp2:1996193} {\bibfield  {journal} {\bibinfo
  {journal} {J. Phys. II}\ }\textbf {\bibinfo {volume} {6}},\ \bibinfo {pages}
  {477--491} (\bibinfo {year} {1996})}\BibitemShut {NoStop}%
\bibitem [{\citenamefont {Gagliardi}\ \emph {et~al.}(2007)\citenamefont
  {Gagliardi}, \citenamefont {Castells}, \citenamefont {R{\`{a}}fols},
  \citenamefont {Ros{\'{e}}s},\ and\ \citenamefont {Bosch}}]{Gagliardi2007}%
  \BibitemOpen
  \bibfield  {author} {\bibinfo {author} {\bibfnamefont {Leonardo~G.}\
  \bibnamefont {Gagliardi}}, \bibinfo {author} {\bibfnamefont {Cecilia~B.}\
  \bibnamefont {Castells}}, \bibinfo {author} {\bibfnamefont {Clara}\
  \bibnamefont {R{\`{a}}fols}}, \bibinfo {author} {\bibfnamefont {Mart{\'{i}}}\
  \bibnamefont {Ros{\'{e}}s}}, \ and\ \bibinfo {author} {\bibfnamefont
  {Elisabeth}\ \bibnamefont {Bosch}},\ }\bibfield  {title} {\enquote {\bibinfo
  {title} {{Static Dielectric Constants of Acetonitrile/Water Mixtures at
  Different Temperatures and Debye−H{\"{u}}ckel A and a 0 B Parameters for
  Activity Coefficients}},}\ }\href {\doibase 10.1021/je700055p} {\bibfield
  {journal} {\bibinfo  {journal} {J. Chem. Eng. Data}\ }\textbf {\bibinfo
  {volume} {52}},\ \bibinfo {pages} {1103--1107} (\bibinfo {year}
  {2007})}\BibitemShut {NoStop}%
\bibitem [{\citenamefont {Riniker}\ \emph {et~al.}(2012)\citenamefont
  {Riniker}, \citenamefont {Horta}, \citenamefont {Thijssen}, \citenamefont
  {Gupta}, \citenamefont {van Gunsteren},\ and\ \citenamefont
  {H{\"{u}}nenberger}}]{Riniker2012}%
  \BibitemOpen
  \bibfield  {author} {\bibinfo {author} {\bibfnamefont {Sereina}\ \bibnamefont
  {Riniker}}, \bibinfo {author} {\bibfnamefont {Bruno A.~C.}\ \bibnamefont
  {Horta}}, \bibinfo {author} {\bibfnamefont {Bram}\ \bibnamefont {Thijssen}},
  \bibinfo {author} {\bibfnamefont {Saumya}\ \bibnamefont {Gupta}}, \bibinfo
  {author} {\bibfnamefont {Wilfred~F.}\ \bibnamefont {van Gunsteren}}, \ and\
  \bibinfo {author} {\bibfnamefont {Philippe~H.}\ \bibnamefont
  {H{\"{u}}nenberger}},\ }\bibfield  {title} {\enquote {\bibinfo {title}
  {{Temperature Dependence of the Dielectric Permittivity of Acetic Acid,
  Propionic Acid and Their Methyl Esters: A Molecular Dynamics Simulation
  Study}},}\ }\href {\doibase 10.1002/cphc.201100949} {\bibfield  {journal}
  {\bibinfo  {journal} {ChemPhysChem}\ }\textbf {\bibinfo {volume} {13}},\
  \bibinfo {pages} {1182--1190} (\bibinfo {year} {2012})}\BibitemShut {NoStop}%
\bibitem [{\citenamefont {Orhan}(2014)}]{Orhan2014}%
  \BibitemOpen
  \bibfield  {author} {\bibinfo {author} {\bibfnamefont {Mehmet}\ \bibnamefont
  {Orhan}},\ }\bibfield  {title} {\enquote {\bibinfo {title} {{Dielectric and
  Transport Properties of Acetonitrile at Varying Temperatures: a Molecular
  Dynamics Study}},}\ }\href {\doibase 10.5012/bkcs.2014.35.5.1469} {\bibfield
  {journal} {\bibinfo  {journal} {Bull. Korean Chem. Soc.}\ }\textbf {\bibinfo
  {volume} {35}},\ \bibinfo {pages} {1469--1478} (\bibinfo {year}
  {2014})}\BibitemShut {NoStop}%
\bibitem [{\citenamefont {Gongadze}\ and\ \citenamefont
  {Igli{\v{c}}}(2012)}]{Gongadze2012}%
  \BibitemOpen
  \bibfield  {author} {\bibinfo {author} {\bibfnamefont {Ekaterina}\
  \bibnamefont {Gongadze}}\ and\ \bibinfo {author} {\bibfnamefont {Ale{\v{s}}}\
  \bibnamefont {Igli{\v{c}}}},\ }\bibfield  {title} {\enquote {\bibinfo {title}
  {{Decrease of permittivity of an electrolyte solution near a charged surface
  due to saturation and excluded volume effects}},}\ }\href {\doibase
  10.1016/j.bioelechem.2011.12.001} {\bibfield  {journal} {\bibinfo  {journal}
  {Bioelectrochemistry}\ }\textbf {\bibinfo {volume} {87}},\ \bibinfo {pages}
  {199--203} (\bibinfo {year} {2012})}\BibitemShut {NoStop}%
\bibitem [{\citenamefont {Fedorov}\ and\ \citenamefont
  {Kornyshev}(2008{\natexlab{a}})}]{Fedorov2008b}%
  \BibitemOpen
  \bibfield  {author} {\bibinfo {author} {\bibfnamefont {M.}~\bibnamefont
  {Fedorov}}\ and\ \bibinfo {author} {\bibfnamefont {Alexei~A.}\ \bibnamefont
  {Kornyshev}},\ }\bibfield  {title} {\enquote {\bibinfo {title} {{Towards
  understanding the structure and capacitance of electrical double layer in
  ionic liquids}},}\ }\href {\doibase 10.1016/j.electacta.2008.02.065}
  {\bibfield  {journal} {\bibinfo  {journal} {Electrochimica Acta}\ }\textbf
  {\bibinfo {volume} {53}},\ \bibinfo {pages} {6835--6840} (\bibinfo {year}
  {2008}{\natexlab{a}})}\BibitemShut {NoStop}%
\bibitem [{\citenamefont {Fedorov}\ \emph {et~al.}(2010)\citenamefont
  {Fedorov}, \citenamefont {Georgi},\ and\ \citenamefont
  {Kornyshev}}]{Fedorov2010}%
  \BibitemOpen
  \bibfield  {author} {\bibinfo {author} {\bibfnamefont {M.~V.}\ \bibnamefont
  {Fedorov}}, \bibinfo {author} {\bibfnamefont {N.}~\bibnamefont {Georgi}}, \
  and\ \bibinfo {author} {\bibfnamefont {A.~A.}\ \bibnamefont {Kornyshev}},\
  }\bibfield  {title} {\enquote {\bibinfo {title} {{Double layer in ionic
  liquids: The nature of the camel shape of capacitance}},}\ }\href {\doibase
  10.1016/j.elecom.2009.12.019} {\bibfield  {journal} {\bibinfo  {journal}
  {Electrochemistry Communications}\ }\textbf {\bibinfo {volume} {12}},\
  \bibinfo {pages} {296--299} (\bibinfo {year} {2010})}\BibitemShut {NoStop}%
\bibitem [{\citenamefont {Girotto}\ \emph {et~al.}(2017)\citenamefont
  {Girotto}, \citenamefont {dos Santos},\ and\ \citenamefont
  {Levin}}]{Girotto2017}%
  \BibitemOpen
  \bibfield  {author} {\bibinfo {author} {\bibfnamefont {Matheus}\ \bibnamefont
  {Girotto}}, \bibinfo {author} {\bibfnamefont {Alexandre~P.}\ \bibnamefont
  {dos Santos}}, \ and\ \bibinfo {author} {\bibfnamefont {Yan}\ \bibnamefont
  {Levin}},\ }\bibfield  {title} {\enquote {\bibinfo {title} {Simulations of
  ionic liquids confined by metal electrodes using periodic {Green}
  functions},}\ }\href {\doibase 10.1063/1.4989388} {\bibfield  {journal}
  {\bibinfo  {journal} {J. Chem. Phys.}\ }\textbf {\bibinfo {volume} {147}},\
  \bibinfo {pages} {074109} (\bibinfo {year} {2017})}\BibitemShut {NoStop}%
\bibitem [{\citenamefont {Pousaneh}\ \emph {et~al.}(2012)\citenamefont
  {Pousaneh}, \citenamefont {Ciach},\ and\ \citenamefont
  {Macio{\l}ek}}]{Pousaneh2012a}%
  \BibitemOpen
  \bibfield  {author} {\bibinfo {author} {\bibfnamefont {Faezeh}\ \bibnamefont
  {Pousaneh}}, \bibinfo {author} {\bibfnamefont {Alina}\ \bibnamefont {Ciach}},
  \ and\ \bibinfo {author} {\bibfnamefont {Anna}\ \bibnamefont {Macio{\l}ek}},\
  }\bibfield  {title} {\enquote {\bibinfo {title} {{Effect of ions on confined
  near-critical binary aqueous mixture}},}\ }\href {\doibase
  10.1039/C2SM25461A} {\bibfield  {journal} {\bibinfo  {journal} {Soft Matter}\
  }\textbf {\bibinfo {volume} {8}},\ \bibinfo {pages} {7567--7581} (\bibinfo
  {year} {2012})}\BibitemShut {NoStop}%
\bibitem [{\citenamefont {Pousaneh}\ \emph {et~al.}(2014)\citenamefont
  {Pousaneh}, \citenamefont {Ciach},\ and\ \citenamefont
  {Macio{\l}ek}}]{Pousaneh2014}%
  \BibitemOpen
  \bibfield  {author} {\bibinfo {author} {\bibfnamefont {Faezeh}\ \bibnamefont
  {Pousaneh}}, \bibinfo {author} {\bibfnamefont {Alina}\ \bibnamefont {Ciach}},
  \ and\ \bibinfo {author} {\bibfnamefont {Anna}\ \bibnamefont {Macio{\l}ek}},\
  }\bibfield  {title} {\enquote {\bibinfo {title} {{How ions in solution can
  change the sign of the critical Casimir potential}},}\ }\href {\doibase
  10.1039/c3sm51972d} {\bibfield  {journal} {\bibinfo  {journal} {Soft Matter}\
  }\textbf {\bibinfo {volume} {10}},\ \bibinfo {pages} {470--483} (\bibinfo
  {year} {2014})}\BibitemShut {NoStop}%
\bibitem [{\citenamefont {Fedorov}\ and\ \citenamefont
  {Kornyshev}(2008{\natexlab{b}})}]{Fedorov2008a}%
  \BibitemOpen
  \bibfield  {author} {\bibinfo {author} {\bibfnamefont {M.~V.}\ \bibnamefont
  {Fedorov}}\ and\ \bibinfo {author} {\bibfnamefont {A.~A.}\ \bibnamefont
  {Kornyshev}},\ }\bibfield  {title} {\enquote {\bibinfo {title} {{Ionic Liquid
  Near a Charged Wall: Structure and Capacitance of Electrical Double
  Layer}},}\ }\href {\doibase 10.1021/jp803440q} {\bibfield  {journal}
  {\bibinfo  {journal} {The Journal of Physical Chemistry B}\ }\textbf
  {\bibinfo {volume} {112}},\ \bibinfo {pages} {11868--11872} (\bibinfo {year}
  {2008}{\natexlab{b}})}\BibitemShut {NoStop}%
\bibitem [{\citenamefont {Carnahan}\ and\ \citenamefont
  {Starling}(1969)}]{Carnahan1969EquationSpheres}%
  \BibitemOpen
  \bibfield  {author} {\bibinfo {author} {\bibfnamefont {Norman~F.}\
  \bibnamefont {Carnahan}}\ and\ \bibinfo {author} {\bibfnamefont {Kenneth~E.}\
  \bibnamefont {Starling}},\ }\bibfield  {title} {\enquote {\bibinfo {title}
  {{Equation of State for Nonattracting Rigid Spheres}},}\ }\href {\doibase
  10.1063/1.1672048} {\bibfield  {journal} {\bibinfo  {journal} {The Journal of
  Chemical Physics}\ }\textbf {\bibinfo {volume} {51}},\ \bibinfo {pages}
  {635--636} (\bibinfo {year} {1969})}\BibitemShut {NoStop}%
\bibitem [{\citenamefont {Gongadze}\ and\ \citenamefont
  {Igli{\v{c}}}(2015)}]{Gongadze2015}%
  \BibitemOpen
  \bibfield  {author} {\bibinfo {author} {\bibfnamefont {Ekaterina}\
  \bibnamefont {Gongadze}}\ and\ \bibinfo {author} {\bibfnamefont {Ale{\v{s}}}\
  \bibnamefont {Igli{\v{c}}}},\ }\bibfield  {title} {\enquote {\bibinfo {title}
  {{Asymmetric size of ions and orientational ordering of water dipoles in
  electric double layer model - an analytical mean-field approach}},}\ }\href
  {\doibase 10.1016/j.electacta.2015.07.179} {\bibfield  {journal} {\bibinfo
  {journal} {Electrochim. Acta}\ }\textbf {\bibinfo {volume} {178}},\ \bibinfo
  {pages} {541--545} (\bibinfo {year} {2015})}\BibitemShut {NoStop}%
\bibitem [{\citenamefont {Sin}\ \emph {et~al.}(2015)\citenamefont {Sin},
  \citenamefont {Im},\ and\ \citenamefont {Kim}}]{Sin2015}%
  \BibitemOpen
  \bibfield  {author} {\bibinfo {author} {\bibfnamefont {Jun~Sik}\ \bibnamefont
  {Sin}}, \bibinfo {author} {\bibfnamefont {Song~Jin}\ \bibnamefont {Im}}, \
  and\ \bibinfo {author} {\bibfnamefont {Kwang~Il}\ \bibnamefont {Kim}},\
  }\bibfield  {title} {\enquote {\bibinfo {title} {{Asymmetric electrostatic
  properties of an electric double layer: A generalized Poisson-Boltzmann
  approach taking into account non-uniform size effects and water
  polarization}},}\ }\href {\doibase 10.1016/j.electacta.2014.11.119}
  {\bibfield  {journal} {\bibinfo  {journal} {Electrochim. Acta}\ }\textbf
  {\bibinfo {volume} {153}},\ \bibinfo {pages} {531--539} (\bibinfo {year}
  {2015})}\BibitemShut {NoStop}%
\bibitem [{\citenamefont {Gongadze}\ \emph {et~al.}(2018)\citenamefont
  {Gongadze}, \citenamefont {Mesarec}, \citenamefont {Kralj-Iglic},\ and\
  \citenamefont {Iglic}}]{Gongadze2018}%
  \BibitemOpen
  \bibfield  {author} {\bibinfo {author} {\bibfnamefont {Ekaterina}\
  \bibnamefont {Gongadze}}, \bibinfo {author} {\bibfnamefont {Luka}\
  \bibnamefont {Mesarec}}, \bibinfo {author} {\bibfnamefont {Veronika}\
  \bibnamefont {Kralj-Iglic}}, \ and\ \bibinfo {author} {\bibfnamefont {Ales}\
  \bibnamefont {Iglic}},\ }\bibfield  {title} {\enquote {\bibinfo {title}
  {{Asymmetric Finite Size of Ions and Orientational Ordering of Water in
  Electric Double Layer Theory Within Lattice Model}},}\ }\href {\doibase
  10.2174/1389557518666180626111927} {\bibfield  {journal} {\bibinfo  {journal}
  {Mini-Reviews Med. Chem.}\ }\textbf {\bibinfo {volume} {18}},\ \bibinfo
  {pages} {1559--1566} (\bibinfo {year} {2018})}\BibitemShut {NoStop}%
\bibitem [{\citenamefont {Lockett}\ \emph
  {et~al.}(2008{\natexlab{b}})\citenamefont {Lockett}, \citenamefont {Sedev},
  \citenamefont {Ralston}, \citenamefont {Horne},\ and\ \citenamefont
  {Rodopoulos}}]{Lockett2008c}%
  \BibitemOpen
  \bibfield  {author} {\bibinfo {author} {\bibfnamefont {Vera}\ \bibnamefont
  {Lockett}}, \bibinfo {author} {\bibfnamefont {Rossen}\ \bibnamefont {Sedev}},
  \bibinfo {author} {\bibfnamefont {John}\ \bibnamefont {Ralston}}, \bibinfo
  {author} {\bibfnamefont {Mike}\ \bibnamefont {Horne}}, \ and\ \bibinfo
  {author} {\bibfnamefont {Theo}\ \bibnamefont {Rodopoulos}},\ }\bibfield
  {title} {\enquote {\bibinfo {title} {{Differential Capacitance of the
  Electrical Double Layer in Imidazolium-Based Ionic Liquids: Influence of
  Potential, Cation Size, and Temperature}},}\ }\href {\doibase
  10.1021/jp7100732} {\bibfield  {journal} {\bibinfo  {journal} {J. Phys. Chem.
  C}\ }\textbf {\bibinfo {volume} {112}},\ \bibinfo {pages} {7486--7495}
  (\bibinfo {year} {2008}{\natexlab{b}})}\BibitemShut {NoStop}%
\bibitem [{\citenamefont {Lockett}\ \emph {et~al.}(2010)\citenamefont
  {Lockett}, \citenamefont {Horne}, \citenamefont {Sedev}, \citenamefont
  {Rodopoulos},\ and\ \citenamefont {Ralston}}]{Lockett2010}%
  \BibitemOpen
  \bibfield  {author} {\bibinfo {author} {\bibfnamefont {Vera}\ \bibnamefont
  {Lockett}}, \bibinfo {author} {\bibfnamefont {Mike}\ \bibnamefont {Horne}},
  \bibinfo {author} {\bibfnamefont {Rossen}\ \bibnamefont {Sedev}}, \bibinfo
  {author} {\bibfnamefont {Theo}\ \bibnamefont {Rodopoulos}}, \ and\ \bibinfo
  {author} {\bibfnamefont {John}\ \bibnamefont {Ralston}},\ }\bibfield  {title}
  {\enquote {\bibinfo {title} {{Differential capacitance of the double layer at
  the electrode/ionic liquids interface}},}\ }\href {\doibase
  10.1039/c0cp00170h} {\bibfield  {journal} {\bibinfo  {journal} {Phys. Chem.
  Chem. Phys.}\ }\textbf {\bibinfo {volume} {12}},\ \bibinfo {pages} {12499}
  (\bibinfo {year} {2010})}\BibitemShut {NoStop}%
\bibitem [{\citenamefont {Butka}\ \emph {et~al.}(2008)\citenamefont {Butka},
  \citenamefont {Vale}, \citenamefont {Saracsan}, \citenamefont {Rybarsch},
  \citenamefont {Weiss},\ and\ \citenamefont
  {Schröer}}]{butka:08:ILPhaseTransitions}%
  \BibitemOpen
  \bibfield  {author} {\bibinfo {author} {\bibfnamefont {Annamaria}\
  \bibnamefont {Butka}}, \bibinfo {author} {\bibfnamefont {Vlad~Romeo}\
  \bibnamefont {Vale}}, \bibinfo {author} {\bibfnamefont {Dragos}\ \bibnamefont
  {Saracsan}}, \bibinfo {author} {\bibfnamefont {Cornelia}\ \bibnamefont
  {Rybarsch}}, \bibinfo {author} {\bibfnamefont {Volker~C.}\ \bibnamefont
  {Weiss}}, \ and\ \bibinfo {author} {\bibfnamefont {Wolffram}\ \bibnamefont
  {Schröer}},\ }\bibfield  {title} {\enquote {\bibinfo {title} {Liquid-liquid
  phase transition in solutions of ionic liquids with halide anions:
  Criticality and corresponding states},}\ }\href {\doibase
  10.1351/pac200880071613} {\bibfield  {journal} {\bibinfo  {journal} {Pure
  Appl. Chem.}\ }\textbf {\bibinfo {volume} {80}},\ \bibinfo {pages}
  {1613--1630} (\bibinfo {year} {2008})}\BibitemShut {NoStop}%
\bibitem [{\citenamefont {Crosthwaite}\ \emph {et~al.}(2004)\citenamefont
  {Crosthwaite}, \citenamefont {Aki}, \citenamefont {Maginn},\ and\
  \citenamefont {Brennecke}}]{Crosthwaite2004}%
  \BibitemOpen
  \bibfield  {author} {\bibinfo {author} {\bibfnamefont {Jacob~M.}\
  \bibnamefont {Crosthwaite}}, \bibinfo {author} {\bibfnamefont {Sudhir N.
  V.~K.}\ \bibnamefont {Aki}}, \bibinfo {author} {\bibfnamefont {Edward~J.}\
  \bibnamefont {Maginn}}, \ and\ \bibinfo {author} {\bibfnamefont {Joan~F.}\
  \bibnamefont {Brennecke}},\ }\bibfield  {title} {\enquote {\bibinfo {title}
  {Liquid phase behavior of imidazolium-based ionic liquids with alcohols},}\
  }\href {\doibase 10.1021/jp037774x} {\bibfield  {journal} {\bibinfo
  {journal} {J. Phys. Chem. B}\ }\textbf {\bibinfo {volume} {108}},\ \bibinfo
  {pages} {5113--5119} (\bibinfo {year} {2004})}\BibitemShut {NoStop}%
\bibitem [{\citenamefont {Rotrekl}\ \emph {et~al.}(2017)\citenamefont
  {Rotrekl}, \citenamefont {Storch}, \citenamefont {Vel{\'{\i}}{\v{s}}ek},
  \citenamefont {Schröer}, \citenamefont {Jacquemin}, \citenamefont {Wagner},
  \citenamefont {Husson},\ and\ \citenamefont
  {Bendov{\'{a}}}}]{rotrekl_bendova:17:ILSDemixing}%
  \BibitemOpen
  \bibfield  {author} {\bibinfo {author} {\bibfnamefont {Jan}\ \bibnamefont
  {Rotrekl}}, \bibinfo {author} {\bibfnamefont {Jan}\ \bibnamefont {Storch}},
  \bibinfo {author} {\bibfnamefont {Petr}\ \bibnamefont
  {Vel{\'{\i}}{\v{s}}ek}}, \bibinfo {author} {\bibfnamefont {Wolffram}\
  \bibnamefont {Schröer}}, \bibinfo {author} {\bibfnamefont {Johan}\
  \bibnamefont {Jacquemin}}, \bibinfo {author} {\bibfnamefont {Zden{\v{e}}k}\
  \bibnamefont {Wagner}}, \bibinfo {author} {\bibfnamefont {Pascale}\
  \bibnamefont {Husson}}, \ and\ \bibinfo {author} {\bibfnamefont {Magdalena}\
  \bibnamefont {Bendov{\'{a}}}},\ }\bibfield  {title} {\enquote {\bibinfo
  {title} {Liquid phase behavior in systems of 1-butyl-3-alkylimidazolium
  bis$\lbrace$(trifluoromethyl)sulfonyl$\rbrace$imide ionic liquids with water:
  Influence of the structure of the c5 alkyl substituent},}\ }\href {\doibase
  10.1007/s10953-017-0659-y} {\bibfield  {journal} {\bibinfo  {journal} {J.
  Solution Chem.}\ }\textbf {\bibinfo {volume} {46}},\ \bibinfo {pages}
  {1456--1474} (\bibinfo {year} {2017})}\BibitemShut {NoStop}%
\bibitem [{\citenamefont {Alam}\ \emph
  {et~al.}(2008{\natexlab{a}})\citenamefont {Alam}, \citenamefont {Islam},
  \citenamefont {Okajima},\ and\ \citenamefont {Ohsaka}}]{Alam2008a}%
  \BibitemOpen
  \bibfield  {author} {\bibinfo {author} {\bibfnamefont {Muhammad~Tanzirul}\
  \bibnamefont {Alam}}, \bibinfo {author} {\bibfnamefont {Md.~Mominul}\
  \bibnamefont {Islam}}, \bibinfo {author} {\bibfnamefont {Takeyoshi}\
  \bibnamefont {Okajima}}, \ and\ \bibinfo {author} {\bibfnamefont {Takeo}\
  \bibnamefont {Ohsaka}},\ }\bibfield  {title} {\enquote {\bibinfo {title}
  {{Capacitance Measurements in a Series of Room-Temperature Ionic Liquids at
  Glassy Carbon and Gold Electrode Interfaces}},}\ }\href {\doibase
  10.1021/jp804620m} {\bibfield  {journal} {\bibinfo  {journal} {The Journal of
  Physical Chemistry C}\ }\textbf {\bibinfo {volume} {112}},\ \bibinfo {pages}
  {16600--16608} (\bibinfo {year} {2008}{\natexlab{a}})}\BibitemShut {NoStop}%
\bibitem [{\citenamefont {Forse}\ \emph {et~al.}(2016)\citenamefont {Forse},
  \citenamefont {Merlet}, \citenamefont {Griffin},\ and\ \citenamefont
  {Grey}}]{forse:jacs:16:chmec}%
  \BibitemOpen
  \bibfield  {author} {\bibinfo {author} {\bibfnamefont {Alexander~C}\
  \bibnamefont {Forse}}, \bibinfo {author} {\bibfnamefont {Céline}\
  \bibnamefont {Merlet}}, \bibinfo {author} {\bibfnamefont {John~M}\
  \bibnamefont {Griffin}}, \ and\ \bibinfo {author} {\bibfnamefont {Clare~P}\
  \bibnamefont {Grey}},\ }\bibfield  {title} {\enquote {\bibinfo {title} {{New
  Perspectives on the Charging Mechanisms of Supercapacitors}},}\ }\href
  {\doibase DOI: 10.1021/jacs.6b02115} {\bibfield  {journal} {\bibinfo
  {journal} {J. Am. Chem. Soc.}\ }\textbf {\bibinfo {volume} {138}},\ \bibinfo
  {pages} {5731--5744} (\bibinfo {year} {2016})}\BibitemShut {NoStop}%
\bibitem [{\citenamefont {Breitsprecher}\ \emph {et~al.}(2017)\citenamefont
  {Breitsprecher}, \citenamefont {Abele}, \citenamefont {Kondrat},\ and\
  \citenamefont {Holm}}]{breitsprecher17a}%
  \BibitemOpen
  \bibfield  {author} {\bibinfo {author} {\bibfnamefont {Konrad}\ \bibnamefont
  {Breitsprecher}}, \bibinfo {author} {\bibfnamefont {Manuel}\ \bibnamefont
  {Abele}}, \bibinfo {author} {\bibfnamefont {Svystolsav}\ \bibnamefont
  {Kondrat}}, \ and\ \bibinfo {author} {\bibfnamefont {Christian}\ \bibnamefont
  {Holm}},\ }\bibfield  {title} {\enquote {\bibinfo {title} {The effect of
  finite pore length on ion structure and charging},}\ }\href {\doibase
  10.1063/1.4986346} {\bibfield  {journal} {\bibinfo  {journal} {The Journal of
  Chemical Physics}\ } (\bibinfo {year} {2017}),\
  10.1063/1.4986346}\BibitemShut {NoStop}%
\bibitem [{\citenamefont {Janssen}\ \emph {et~al.}(2014)\citenamefont
  {Janssen}, \citenamefont {H{\"{a}}rtel},\ and\ \citenamefont {van
  Roij}}]{janssen:14:prl}%
  \BibitemOpen
  \bibfield  {author} {\bibinfo {author} {\bibfnamefont {Mathijs}\ \bibnamefont
  {Janssen}}, \bibinfo {author} {\bibfnamefont {Andreas}\ \bibnamefont
  {H{\"{a}}rtel}}, \ and\ \bibinfo {author} {\bibfnamefont {René}\
  \bibnamefont {van Roij}},\ }\bibfield  {title} {\enquote {\bibinfo {title}
  {{Boosting Capacitive Blue-Energy and Desalination Devices with Waste
  Heat}},}\ }\href {\doibase 10.1103/PhysRevLett.113.268501} {\bibfield
  {journal} {\bibinfo  {journal} {Phys. Rev. Lett.}\ }\textbf {\bibinfo
  {volume} {113}},\ \bibinfo {pages} {268501} (\bibinfo {year}
  {2014})}\BibitemShut {NoStop}%
\bibitem [{\citenamefont {H{\"{a}}rtel}\ \emph {et~al.}(2015)\citenamefont
  {H{\"{a}}rtel}, \citenamefont {Janssen}, \citenamefont {Weingarth},
  \citenamefont {Presser},\ and\ \citenamefont {van Roij}}]{haertel:15:ees}%
  \BibitemOpen
  \bibfield  {author} {\bibinfo {author} {\bibfnamefont {Andreas}\ \bibnamefont
  {H{\"{a}}rtel}}, \bibinfo {author} {\bibfnamefont {Mathijs}\ \bibnamefont
  {Janssen}}, \bibinfo {author} {\bibfnamefont {Daniel}\ \bibnamefont
  {Weingarth}}, \bibinfo {author} {\bibfnamefont {Volker}\ \bibnamefont
  {Presser}}, \ and\ \bibinfo {author} {\bibfnamefont {René}\ \bibnamefont
  {van Roij}},\ }\bibfield  {title} {\enquote {\bibinfo {title}
  {{Heat-to-Current Conversion of Low-Grade Heat from a Thermocapacitive Cycle
  by Supercapacitors}},}\ }\href {\doibase 10.1039/c5ee01192b} {\bibfield
  {journal} {\bibinfo  {journal} {Energ. Environ. Sci.}\ }\textbf {\bibinfo
  {volume} {8}},\ \bibinfo {pages} {2396--2401} (\bibinfo {year}
  {2015})}\BibitemShut {NoStop}%
\bibitem [{\citenamefont {Wang}\ \emph {et~al.}(2015)\citenamefont {Wang},
  \citenamefont {Feng}, \citenamefont {Yang}, \citenamefont {Hau},
  \citenamefont {Munro}, \citenamefont {Ferreira-Yang},\ and\ \citenamefont
  {Chen}}]{wang:15:nanolett}%
  \BibitemOpen
  \bibfield  {author} {\bibinfo {author} {\bibfnamefont {Jianjian}\
  \bibnamefont {Wang}}, \bibinfo {author} {\bibfnamefont {Shien-Ping}\
  \bibnamefont {Feng}}, \bibinfo {author} {\bibfnamefont {Yuan}\ \bibnamefont
  {Yang}}, \bibinfo {author} {\bibfnamefont {Nga~Yu}\ \bibnamefont {Hau}},
  \bibinfo {author} {\bibfnamefont {Mary}\ \bibnamefont {Munro}}, \bibinfo
  {author} {\bibfnamefont {Emerald}\ \bibnamefont {Ferreira-Yang}}, \ and\
  \bibinfo {author} {\bibfnamefont {Gang}\ \bibnamefont {Chen}},\ }\bibfield
  {title} {\enquote {\bibinfo {title} {{"Thermal Charging" Phenomenon in
  Electrical Double Layer Capacitors}},}\ }\href {\doibase
  10.1021/acs.nanolett.5b01761} {\bibfield  {journal} {\bibinfo  {journal}
  {Nano Lett.}\ }\textbf {\bibinfo {volume} {15}},\ \bibinfo {pages}
  {5784--5790} (\bibinfo {year} {2015})}\BibitemShut {NoStop}%
\bibitem [{\citenamefont {Janssen\;}\ and\ \citenamefont {van
  Roij}(2017)}]{Janssen2017}%
  \BibitemOpen
  \bibfield  {author} {\bibinfo {author} {\bibfnamefont {Mathijs}\ \bibnamefont
  {Janssen\;}}\ and\ \bibinfo {author} {\bibfnamefont {Ren\'{e}}\ \bibnamefont
  {van Roij}},\ }\bibfield  {title} {\enquote {\bibinfo {title} {Reversible
  heating in electric double layer capacitors},}\ }\href {\doibase
  10.1103/physrevlett.118.096001} {\bibfield  {journal} {\bibinfo  {journal}
  {Phys. Rev. Lett.}\ }\textbf {\bibinfo {volume} {118}},\ \bibinfo {pages}
  {96001} (\bibinfo {year} {2017})}\BibitemShut {NoStop}%
\bibitem [{\citenamefont {Janssen}\ \emph {et~al.}(2017)\citenamefont
  {Janssen}, \citenamefont {Griffioen}, \citenamefont {Biesheuvel},
  \citenamefont {van Roij},\ and\ \citenamefont {Ern{\'{e}}}}]{Janssen2017a}%
  \BibitemOpen
  \bibfield  {author} {\bibinfo {author} {\bibfnamefont {Mathijs}\ \bibnamefont
  {Janssen}}, \bibinfo {author} {\bibfnamefont {Elian}\ \bibnamefont
  {Griffioen}}, \bibinfo {author} {\bibfnamefont {P.{\hspace{0.167em}}M.}\
  \bibnamefont {Biesheuvel}}, \bibinfo {author} {\bibfnamefont {Ren{\'{e}}}\
  \bibnamefont {van Roij}}, \ and\ \bibinfo {author} {\bibfnamefont {Ben}\
  \bibnamefont {Ern{\'{e}}}},\ }\bibfield  {title} {\enquote {\bibinfo {title}
  {{Coulometry and Calorimetry of Electric Double Layer Formation in Porous
  Electrodes}},}\ }\href {\doibase 10.1103/physrevlett.119.166002} {\bibfield
  {journal} {\bibinfo  {journal} {Phys. Rev. Lett.}\ }\textbf {\bibinfo
  {volume} {119}},\ \bibinfo {pages} {166002} (\bibinfo {year}
  {2017})}\BibitemShut {NoStop}%
\bibitem [{\citenamefont {Alam}\ \emph
  {et~al.}(2008{\natexlab{b}})\citenamefont {Alam}, \citenamefont {Islam},
  \citenamefont {Okajima},\ and\ \citenamefont {Ohsaka}}]{Alam2008}%
  \BibitemOpen
  \bibfield  {author} {\bibinfo {author} {\bibfnamefont {M.~T.}\ \bibnamefont
  {Alam}}, \bibinfo {author} {\bibfnamefont {M.~M.}\ \bibnamefont {Islam}},
  \bibinfo {author} {\bibfnamefont {T.}~\bibnamefont {Okajima}}, \ and\
  \bibinfo {author} {\bibfnamefont {T.}~\bibnamefont {Ohsaka}},\ }\bibfield
  {title} {\enquote {\bibinfo {title} {Capacitance measurements in a series of
  room-temperature ionic liquids at glassy carbon and gold electrode
  interfaces},}\ }\href@noop {} {\bibfield  {journal} {\bibinfo  {journal} {J.
  Phys. Chem. C}\ }\textbf {\bibinfo {volume} {112}},\ \bibinfo {pages}
  {16600--16608} (\bibinfo {year} {2008}{\natexlab{b}})}\BibitemShut {NoStop}%
\bibitem [{\citenamefont {Sha}\ \emph {et~al.}(2014)\citenamefont {Sha},
  \citenamefont {Dou}, \citenamefont {Luo}, \citenamefont {Zhu},\ and\
  \citenamefont {Wu}}]{Sha2014}%
  \BibitemOpen
  \bibfield  {author} {\bibinfo {author} {\bibfnamefont {Maolin}\ \bibnamefont
  {Sha}}, \bibinfo {author} {\bibfnamefont {Qiang}\ \bibnamefont {Dou}},
  \bibinfo {author} {\bibfnamefont {Fabao}\ \bibnamefont {Luo}}, \bibinfo
  {author} {\bibfnamefont {Guanglai}\ \bibnamefont {Zhu}}, \ and\ \bibinfo
  {author} {\bibfnamefont {Guozhong}\ \bibnamefont {Wu}},\ }\bibfield  {title}
  {\enquote {\bibinfo {title} {Molecular insights into the electric double
  layers of ionic liquids on au(100) electrodes},}\ }\href {\doibase
  10.1021/am502413m} {\bibfield  {journal} {\bibinfo  {journal} {{ACS} Applied
  Materials {\&} Interfaces}\ }\textbf {\bibinfo {volume} {6}},\ \bibinfo
  {pages} {12556--12565} (\bibinfo {year} {2014})}\BibitemShut {NoStop}%
\end{thebibliography}%

\end{document}